\documentclass[twocol]{ametsocV6}
\usepackage{moreverb}
\usepackage{graphicx,overpic,natbib}
\usepackage[skip=2pt]{caption}
\usepackage{bm,url,color,colortbl,xcolor,soul}
\usepackage{url}
\usepackage{multirow}
\usepackage{booktabs} %for table
\usepackage{float}
\usepackage{wrapfig}
\usepackage{amsmath}
\newcommand\BibTeX{{\rmfamily B\kern-.05em \textsc{i\kern-.025em b}\kern-.08em
T\kern-.1667em\lower.7ex\hbox{E}\kern-.125emX}}
\newcommand{\Fig}[1]{Figure~\ref{#1}}
\newcommand{\Tab}[1]{Table~\ref{#1}}

\newcommand{\EQ}{\begin{equation}}
\newcommand{\EN}{\end{equation}}
\newcommand{\Eq}[1]{Equation~(\ref{#1})}

\newcommand{\febm}{0228\_\rm{NA1\_\bar{\kappa}_{org}}}
\newcommand{\febu}{0228\_\rm{NA1\_\kappa_{org}^{max}}}

\graphicspath{{./fig/}{./png/}}

\bibpunct{(}{)}{;}{a}{}{,}

\title{Large-eddy simulations of marine boundary-layer clouds associated with cold air outbreak during the ACTIVATE campaign. Part II: aerosol--meteorology--cloud interaction}
\authors{   
Xiang-Yu Li\aff{a}\correspondingauthor{Xiang-Yu Li, xiangyu.li@pnnl.gov},
Hailong Wang\aff{a}\correspondingauthor{Hailong Wang, hailong.wang@pnnl.gov},
Jingyi Chen\aff{a},
Satoshi Endo\aff{b},
Simon Kirschler\aff{c},
Christiane Voigt\aff{c},
Ewan Crosbie\aff{d,e},
Luke D Ziemba\aff{d},
David Painemal\aff{d,e}, 
Brian Cairns\aff{f},
Johnathan W Hair\aff{d},
Andrea F. Corral\aff{g}, 
Claire Robinson\aff{e},
Hossein Dadashazar \aff{g},
Armin Sorooshian\aff{g, h},
Gao Chen\aff{d},
Richard Anthony Ferrare\aff{d},
Mary M Kleb\aff{d}, 
Hongyu Liu\aff{i},
Richard Moore\aff{d},
Amy Jo Scarino\aff{e,d},
Michael A. Shook\aff{d},
Taylor J Shingler\aff{d},
Kenneth Lee Thornhill\aff{e,d},
Florian Tornow\aff{f,j},
Heng Xiao\aff{a},
Xubin Zeng\aff{h}
}
\affiliation{
\aff{a}Pacific Northwest National Laboratory \\
\aff{b}Brookhaven National Laboratory, Upton, NY, United States \\
\aff{c}Institut f{\"u}r Physik der Atmosphäre, Deutsches Zentrum für Luft- und Raumfahrt (DLR), Oberpfaffenhofen, Germany, and Institute for Physics of the Atmosphere, Johannes Gutenberg-University Mainz, Germany \\
\aff{d}NASA Langley Research Center, Hampton, VA, United States \\
\aff{e}Science Systems and Applications, Inc. Hampton, Hampton, VA, United States \\
\aff{f}NASA Goddard Institute for Space Studies, New York, NY, United States \\
\aff{g}University of Arizona, Department of Chemical
and Environmental Engineering, Tucson, AZ, United States \\
\aff{h}University of Arizona, Department of Hydrology and Atmospheric Sciences, Tucson, AZ, United States \\
\aff{i}National Institute of Aerospace, Hampton, VA, United States \\
\aff{j}Columbia University of New York, Center for Climate Systems Research,
Earth Institute, New York, NY, United States
}

\abstract{Aerosol effects on micro-/macro-physical properties of marine stratocumulus clouds
over the Western North Atlantic Ocean (WNAO) are investigated using in-situ
measurements and large-eddy simulations (LES) for two cold air outbreak (CAO)
cases (February 28 and March 1, 2020) during the Aerosol Cloud meTeorology Interactions oVer
the western ATlantic Experiment (ACTIVATE).
The LES is able to reproduce the vertical profiles
of liquid water content (LWC), effective radius $r_{\rm eff}$
and the cloud droplet number concentration $N_c$
from fast cloud droplet probe (FCDP) in-situ measurements
for both cases.
Furthermore, we show that aerosols affect cloud properties
($N_c$, $r_{\rm eff}$, and LWC) via the prescribed bulk hygroscopicity of aerosols ($\bar{\kappa}$) and
aerosol size distributions characteristics. $N_c$, $r_{\rm eff}$,
and liquid water path (LWP) are positively correlated to $\bar{\kappa}$ and aerosol number concentration ($N_a$) while
cloud fractional cover (CFC) is insensitive to $\bar{\kappa}$ and aerosol size distributions for the two cases.
The changes to aerosol size distribution (number concentration, width, and the geometrical diameter) allow
us to disentangle aerosol effects on cloud properties from the
meteorological effects. 
We also use the LES results to evaluate cloud properties from two reanalysis
products, ERA5 and MERRA-2. Comparing to LES, the ERA5 reanalysis is able to capture the time
evolution of LWP and total cloud coverage within the study domain during both CAO
cases while MERRA-2 underestimates them.
}

\begin{document}
\maketitle

\section{Introduction}

Aerosols and clouds pose the largest uncertainty in
climate projection since they mediate the radiative forcing
of the Earth's atmosphere \citep{seinfeld2016improving}.
An increased loading of aerosols in the atmosphere reflects
more incoming solar energy to space and consequentially cools 
the Earth system. The increased aerosol number concentration $N_a$
in an environment with constant liquid water content (LWC)
leads to cloud droplets with smaller size and larger number
concentration $N_c$, which makes the cloud more reflective, often called the first indirect effect, a.k.a. Twomey effect \citep{twomey1977influence}. The reduced droplet sizes can also result in less precipitation and longer cloud lifetime.
The latter effect is known as the second
indirect effect \citep{albrecht1989aerosols}. 
Cloud radiative properties are determined by both the cloud macrophysical properties,
such as the liquid water path (LWP) and cloud fractional coverage (CFC)
and by the cloud microphysical properties, such as $N_c$ and the effecive
radius $r_{\rm eff}$.
Global mean change in radiative forcing due to anthropogenic aerosols, relative to preindustrial era (before year 1850), is roughly estimated to be
between $-1.6$ to $-0.6\, \rm{W\, m}^{-2}$ with a $68\%$ confidence interval in year 2005-2015 \citep{bellouin2020bounding}. However, how the anthropogenic aerosols affect LWC and CFC remains unclear \citep{bellouin2020bounding, mccoy2020untangling},
which is due to the fact that LWC and CFC are predominantly determined by
meteorology states but also modulated by changes in $N_c$ through precipitation
\citep{stevens2009untangling}. This is the so-called aerosol-meteorology-cloud-interaction
(AMCI) problem that encompasses a wide range of spatial-temporal scales from
nm-sized aerosol particles to large-scale atmospheric circulations ($\mathcal{O}(100\, \rm{km})$).
Global Earth system models and even cloud resolving models
suffer from coarse resolutions such that vital cloud macro/micro processes in marine boundary layer
cannot be physically represented.

Here, we focus on the AMCI process in marine boundary-layer stratocumulus,
which is a canonical cloud regime for studying AMCI as this regime is often
in an emergent state of precipitation and is sensitive to AMCI \citep{feingold2010precipitation, seifert2015large}.
In this case, an increased $N_c$ may lead to either an increased
LWC due to the suppression of precipitation \citep{albrecht1989aerosols} or a reduced LWC
due to enhanced entrainment of dry air \citep{ackerman2004impact}.
The overall effects of increasing $N_c$ on LWC depends on the meteorology states
in different cloud regimes.
The suppression of precipitation may alter CFC
by either increasing the cloud lifetime \citep{albrecht1989aerosols} or by influencing the transitioning from open to closed Rayleigh-B\'enard cells of
stratocumulus \citep{rosenfeld2006switching, wang2009modeling}, or by altering mesoscale circulation and lower-level moisture convergence \citep{wang2009modeling2}.
Nevertheless, response of LWP and CFC to $N_c$ is still uncertain in
both the observation and the general circulation models (GCMs).
GCMs typically show a positive correlation between LWP and $N_c$ and
between CFC and $N_c$
due to a direct reduction in the autoconversion of cloud droplets to raindrops by an increased $N_c$, which is parameterized for the collision-coalescence process \citep{ghan2016challenges, bellouin2020bounding}.
Using large-eddy simulations (LES) with prescribed constant $N_c$, \citet{seifert2015large} showed that
the suppression of precipitation due to increased $N_c$ reduces CFC in trade-wind cumulus. However, this effect
is compensated by the Twomey effect and the overall
effect of aerosols on the albedo of cloud
is small. LES with more complete and resolved physical processes, informed and constrained by observations, is needed to unravel discrepancies in GCMs regarding AMCI.

AMCI associated with marine cold-air outbreaks (CAO) is poorly understood and
has been rarely studied due to the more complicated CAO cloud processes and a lack of measurements. 
\citet{Roode2019} performed an LES inter-comparison of
a CAO case to quantify the turbulent transport at length scales between
1 and 10 km and to study the sensitivity of the CAO to $N_c$ and ice microphysics. Using the same microphysics scheme \citep{seifert2006two} and prescribed constant $N_c$, the Dutch Atmospheric LES, the Max-Planck Institute
for Meteorology (MPI) LES, and the PArallelized Large Eddy Simulation Model for Atmospheric and Oceanic Flows (PALM) obtained very
different CFC, LWP, and surface
precipitation in both magnitude and timing. They also show
that a reduction of $N_c$ results in a stronger precipitation, smaller LWP and earlier breakup of clouds, which is consistent
with the findings for subtropical marine stratocumulus in \citet{wang2009modeling}.
As marine CAOs are generally associated with mixed-phase clouds \citep{fletcher2016climatology},
the proper characterization and simulation of ice microphysics
are expected to be critical for AMCI.
Substantial spread among LES models in ice water content (IWC)
due to different parameterizations of ice microphysics 
was observed in \citet{Roode2019}.
\citet{tornow2021preconditioning} examined the role of the riming process
in marine CAO using LES with a prescribed single-mode lognormal aerosol-size distributions and two-moment Morrison cloud microphysics scheme. They showed that
increasing a diagnostic ice nuclei particle concentration intensifies early
and light precipitation, which accelerates the stratocumulus breakup.
Their study showed that only a prognostic aerosol treatment produced plausible cloud regime transitions and indicated that a prognostic INP concentration is needed to capture ice multiplication near the cloud breakup that has often been observed.
However, these LES studies did not explore how
$N_c$ impacted by a realistic setup of several aerosol modes and their hygroscopicity modulates
LWP, CFC, and radiation fluxes during CAO.
Therefore, a prognostic $N_c$, based on model resolved meteorology and measured 
aerosol size distributions and hygroscopicity, is important in understanding AMCI.  

The Western North Atlantic Ocean (WNAO) region is characterized by
a complicated climate system that features weather processes involving
a wide range of spatial-temporal scales \citep{sorooshian2020atmospheric}.
The sea surface temperature (SST) in the WNAO exhibits sharp spatial gradients due
to the Gulf stream, which, together with strong wind, lead to strong
surface heat fluxes within the atmospheric boundary layer \citep{painemaloverview}.
This creates ideal condition for cold air outbreak (CAO) events as investigated
in \citet{Seethala21} and \citet{2021arXiv210706193L}.
Aerosols transported from sources over the continental U.S.,
generated over ocean from shipping emissions and sea spray, and
produced by long-range transport of smoke and dust \citep{corraloverview, aldhaif2020sources, Aldhaif} contribute to the total aerosol number concentration
in WNAO. The AMCI is poorly understood
in this region, which is partly due to limited in-situ measurements of aerosol
and cloud microphyscial processes \citep{sorooshian2019aerosol}.
The Aerosol Cloud meTeorology Interactions oVer the western ATlantic Experiment (ACTIVATE)
campaign aims to unravel AMCI in WNAO by collecting unprecedented in-situ and remote-sensing
statistics of aerosols and cloud properties. To achieve this, the dual-aircraft approach
is being adopted for about 150 flights ($\sim 600$ joint total flight hours) during 2020-2022
in the WNAO region ($25^\circ-50^\circ$N, $60^\circ-85^\circ$W). The lower-flying HU-25 Falcon
(a minimum altitude of 150 m) focuses on measuring in-situ trace gases, aerosol,
cloud properties, thermodynamics, and precipitation. 
The higher-flying King Air (nominal flight altitude of 9 km)
simultaneously acquires remote-sensing retrievals of aerosols and clouds and
deploys dropsondes to measure the meteorological states \citep{sorooshian2019aerosol}.

We aim to study AMCI in the WNAO region by performing
LES, constrained and evaluated by the ACTIVATE measurements.
The measured aerosol size distributions and hygroscopicity, which provide a more realistic cloud condensation nuclei (CCN) pool, are fed
into a two-moment microphysics scheme in the LES, compared to a fixed CCN for the prognostic $N_c$ used in previous LES studies.
To our knowledge, LES investigation of AMCI associated with CAO with realistic aerosol perturbations in the WNAO region has never been done before. 
Furthermore, the ameliorated understanding of AMCI and quantified aerosol-cloud relationships from LES can 
serve as a benchmark to evaluate and improve parameterizations of aerosol-cloud processes in global and regional models. 

Using the same large-scale forcing strategy as in \citet{endo2015racoro},
\citet{2021arXiv210706193L} investigated the marine boundary-layer (BL) and clouds during two CAO events (February 28 and March 1, 2020), which are
characterized by strong temporally and spatially varying meteorological states
in the WNAO region. It focused on the sensitivities of WRF-LES
to large-scale forcing and surface heat fluxes
constrained by the ACTIVATE measurements.
In this companion study of \citet{2021arXiv210706193L}, 
we focus on aerosol effects on clouds under the two CAO conditions.
%, as part of the AMCI during CAO over the WNAO region.
As mentioned above, we use WRF-LES with prognostic $N_c$ based on measured aerosol properties, compare model results with the ACTIVATE cloud measurements, and evaluate cloud macrophysical properties in
reanalysis products for the two CAO events reported in \citet{2021arXiv210706193L}.
The reminder of the paper is organized as follows.
In section~\ref{sec:obs}, we describe ACTIVATE measurements, reanalysis data, and satellite retrievals used in this study.
Section~\ref{sec:numerical} summarizes the WRF-LES numerical experiment
setup. Section~\ref{sec:LES} discusses WRF-LES simulated aerosol effects on CAO clouds in two cases and compared to ACTIVATE measurements. Section~\ref{sec:comp} compares the LES BL structure and clouds with reanalysis and satellite retrievals. We conclude in section~\ref{sec:last}.

\section{Observations, reanalysis, and satellite retrievals}
\label{sec:obs}

\subsection{Aerosol size distribution}
\label{sec:aerosol}

Aerosol particles with diameter $d$ between $3-100\, \rm{nm}$
were measured by the Scanning Mobility Particle Sizer (SMPS, TSI model 3085 differential mobility analyzer and TSI model 3776 condensation particle counter)
and those with sizes larger than $100\, \rm{nm}$ were measured by the
Laser Aerosol Spectrometer (LAS, TSI model 3340) equipped on Falcon HU-25.
The uncertainty for both SMPS and LAS measurements is better than $\pm10\%-20\%$
over the submircron aerosol size range \citep{moore2021sizing}.
The sampling frequency is 1/60 Hz and 1 Hz for SMPS and LAS, respectively.

\begin{figure*}[t!]\begin{center}
\includegraphics[width=0.48\textwidth]{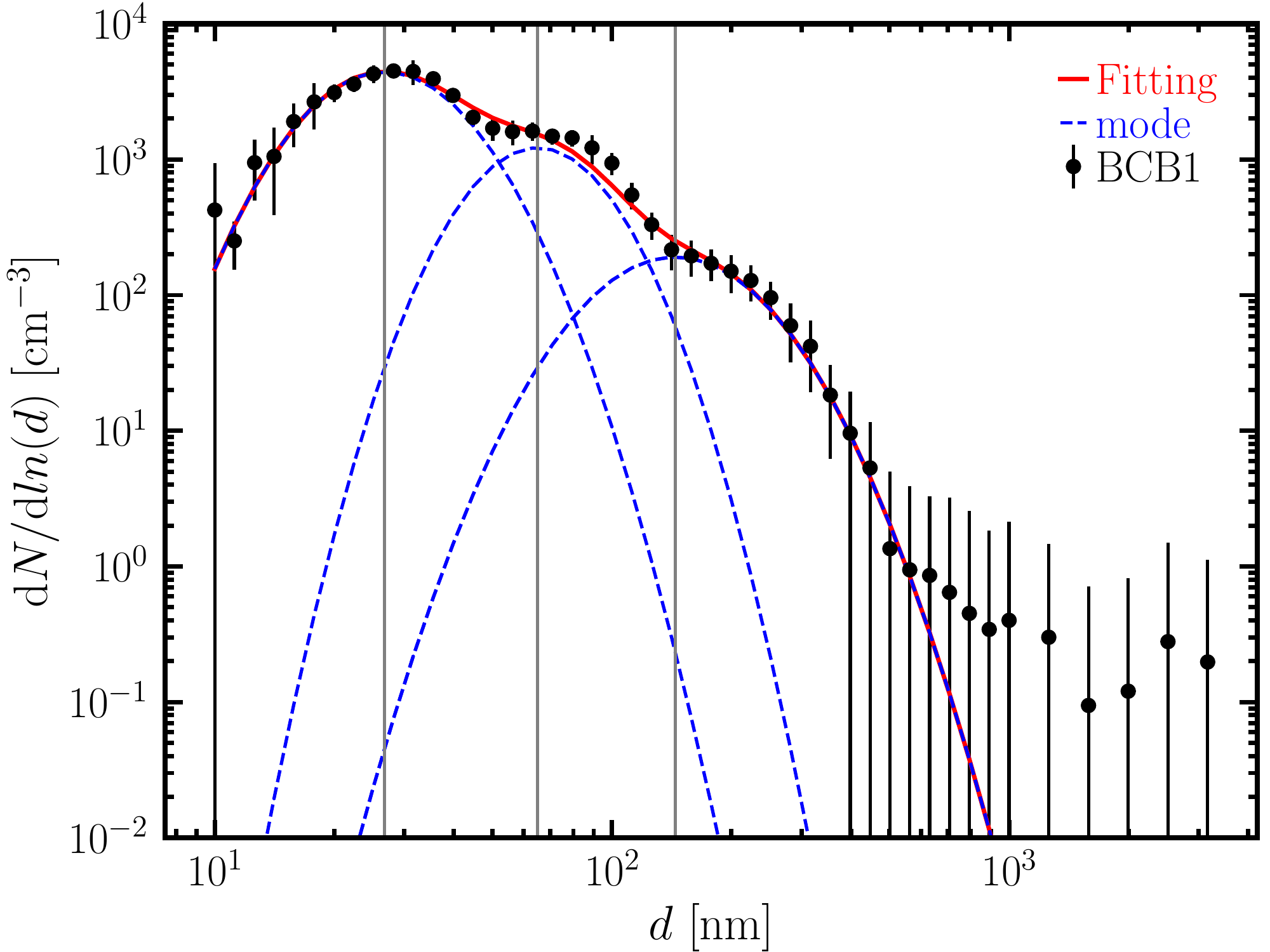}
\includegraphics[width=0.48\textwidth]{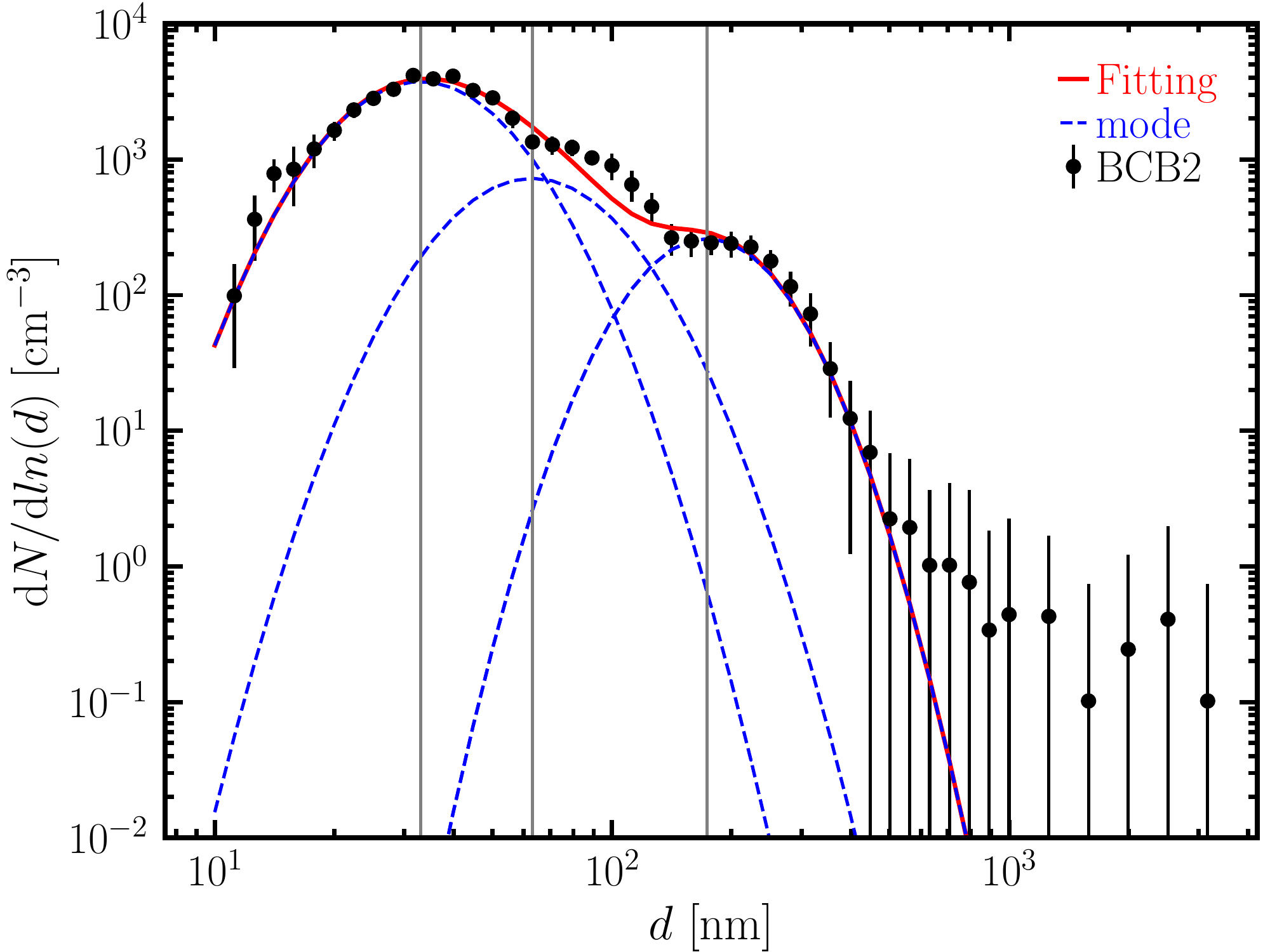}
\end{center}\caption{Aerosol size distributions (black dots) obtained from SMPS
and LAS measurements for the February 28 case during the BCB legs.
The error bars indicate $\pm\sigma$ deviation from the time-averaged aerosol
size distribution during a BCB flight leg.
The red curve represents the final fitted size distribution. The dashed blue curves represent log-normal fitting of individual modes. Fitted parameters are listed in \Tab{tab:fit}. Only particles with $d\ge 10$ nm are used for the fitting.
The coarse mode at $d\ge 1\mu\rm{m}$ is not used for the fitting
because of the very low number concentrations (at least 2 orders of magnitude lower than the accumulation mode) and uncertainties (large error bars).  
}
\label{dNdlnD_60221_60423_3modes_0228}
\end{figure*}

The black dots in \Fig{dNdlnD_60221_60423_3modes_0228} represent the aerosol
size distribution averaged over Below Cloud-Base
(BCB) flight legs, sampled by SMPS and LAS within the study domain (16:00-17:00 UTC) for the February 28 case.
Two BCB flight legs were performed, marked as BCB1 and BCB2, as shown in
\Fig{FCDP_traj}(a) and \Fig{cpc_0228}(a).
%Only particles larger than 10 nm in diameter are shown, considering that uncertainties of measurements
%for smaller particles and impact on the fitting are large.  
Only particles larger than 10 nm in diameter are shown, as smaller nucleation-mode particles are not likely contributing to the population of cloud condensation nuclei at realistic supersaturations.
We fit the measured aerosol size distributions using a log-normal distribution function,
\EQ
\frac{{\rm d}N}{{\rm d}\ln d} = \frac{N^*}{\sqrt{2\pi}\ln \sigma}\exp\left[-\frac{(\ln d -\ln\mu)^2}{2\ln^2\sigma}\right],
\EN
where $N^*$ is the total aerosol number concentration.
$\mu$ and $\sigma$ are the geometric median diameter and
the standard deviation, respectively.
The fitted parameters using three log-normal modes are listed in \Tab{tab:fit}.
The fitting is validated by comparing the number concentration integrated
over the fitted lognormal distribution
%$\bar{N}_{\rm fit}=\int {\rm d}N/{\rm d}\ln d$ to the
$\bar{N}_{\rm fit}=\int {\rm d}N$ to the
measured integrated number concentration
of aerosols $\bar{N}_{\rm a}$ during the BCB flight leg.
The Percent Error (PE) in total particle number for the BCB1 and BCB2
legs are $(\bar{N}_{\rm fit}-\bar{N}_{\rm a})/\bar{N}_{\rm a}\times 100 =-3.7\%$ and $-12.6\%$, respectively.
The relatively low value of PE suggests that the fitting is robust.

\begin{figure*}[t!]\begin{center}
\includegraphics[width=0.48\textwidth]{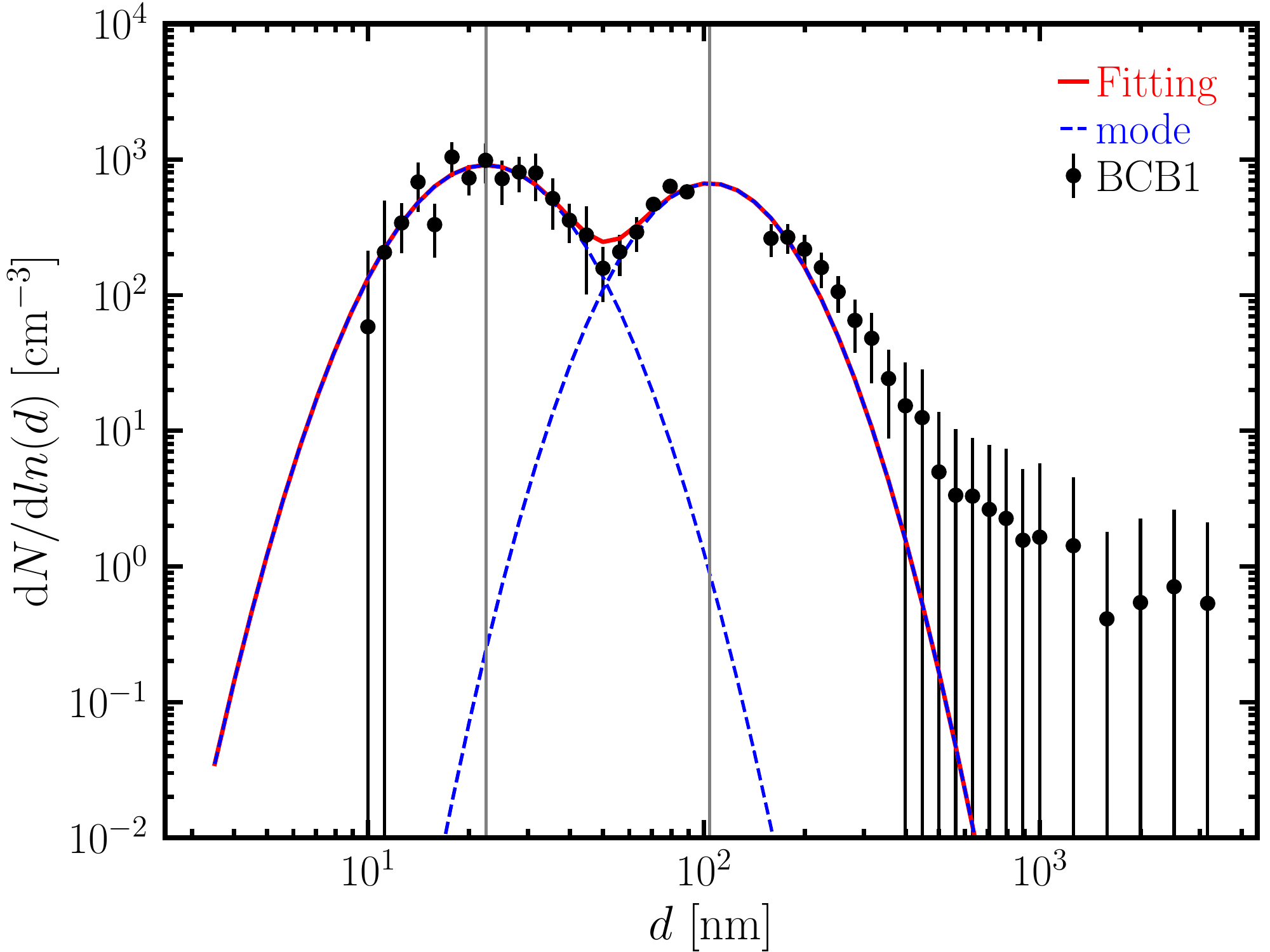}
\includegraphics[width=0.48\textwidth]{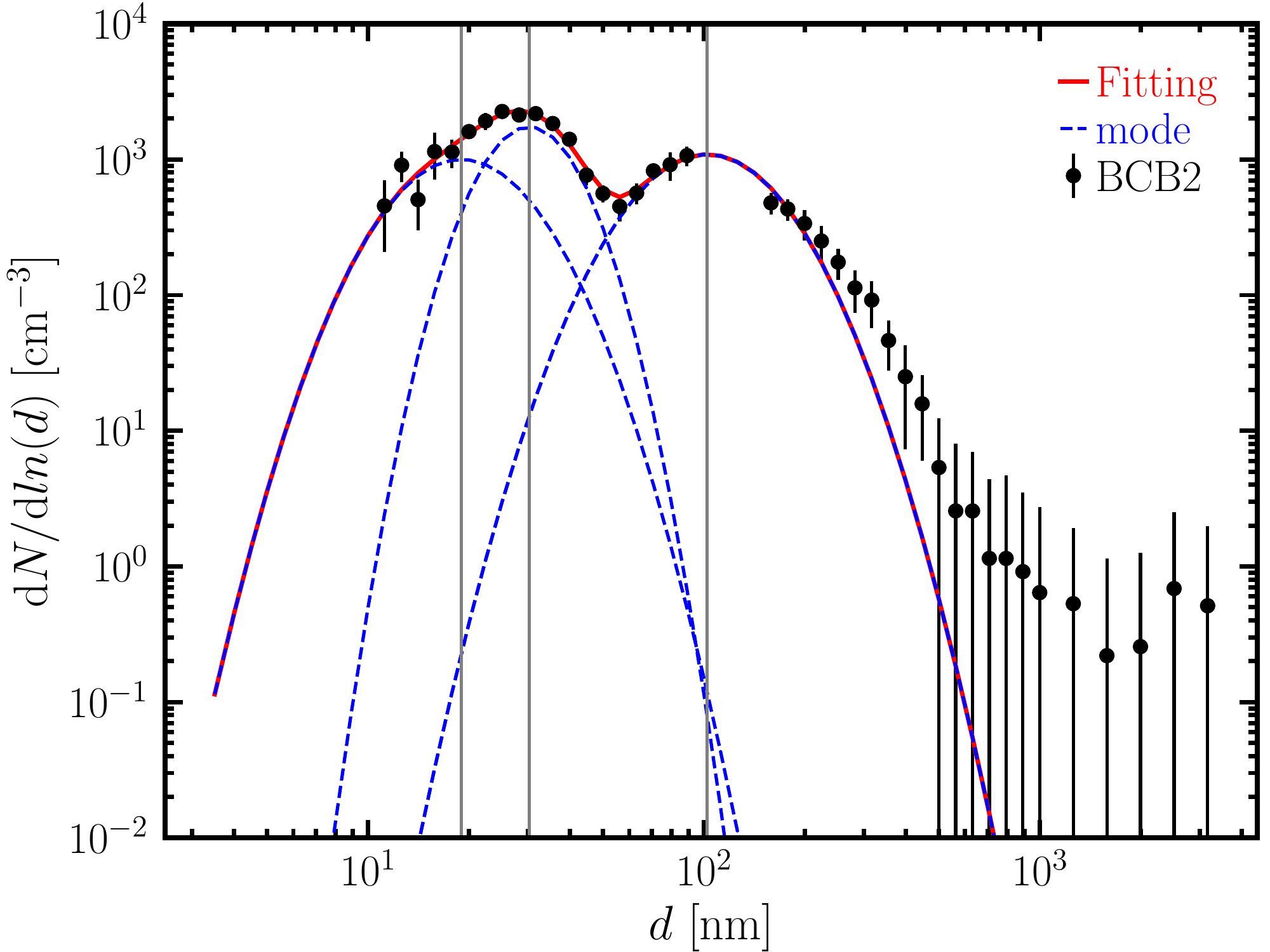}
i%\includegraphics[width=0.32\textwidth]{dNdlnD_57614_57814_3modes_0301}
\end{center}\caption{Same as \Fig{dNdlnD_60221_60423_3modes_0228} but for the March 1 case.
Note that LAS measurement data for the first four bins are missing due
to the failure of the laser power on March 1. Therefore they are not included in the fitting procedure. 
}
\label{dNdlnD_53602_54105_2modes_0301}
\end{figure*}

The aerosol size distribution for the March 1 case is shown
in \Fig{dNdlnD_53602_54105_2modes_0301} and the corresponding
fitted parameters and validation are listed in \Tab{tab:fit0301}.
The same to the February 28 case, two BCB flight legs were performed as shown in
\Fig{FCDP_traj}(b) and \Fig{cpc_0228}(b).
Two and three log-normal modes are used for the fitting for BCB1 and BCB2, respectively.
The fitting is robust as indicated by low values of PEs ($3.1\%$ and $1.3\%$).
It is worth noting that $\bar{N}_a$ from BCB2 is about 2.6 times larger than
that from BCB2, which suggests a substantial spatiotemporal
variation of aerosol particles below cloud-base.
Without interactive aerosol sources and realistic boundary conditions, we do not expect the LES to capture the spatial variation of size distribution and hygroscopicity of aerosols. Thus a uniform size distribution and hygroscopicity is used within the simulation domain.
The same aerosol treatment was used in
\citet{endo2015racoro}, who found that cloud macrophysical properties are insensitive, but cloud microphysical properties are sensitive, to aerosol hygroscopicity. 
However, they estimated the aerosol hygroscopicity according to the K{\"o}hler theory using the aerosol size distribution and CCN concentration. 
We use the direct measurement during the ACTIVATE campaign to ensure a more robust estimation of the bulk hygroscopicity of aerosols $\bar{\kappa}$ as discussed below.

\subsection{Hygroscopicity of aerosol particles}
\label{sec:kappa}

The WRF-LES takes bulk hygroscopicity ($\kappa$) of each size mode, which can be estimated from $\kappa$ and mass of each chemical component.
The bulk $\kappa$ of a solute aerosol particle with mixed component $i$
is given by the simple volume mixing rule \citep{petters2007single}
\EQ
\kappa = \sum_i \epsilon_i \kappa_i,
\label{eq:kappa}
\EN
where $\kappa_i$ is the hygroscopicity parameter of
each individual (dry) component of aerosol particles
and is obtained from Table 1 of \citet{petters2007single} for
both the non-organic and organic components.
The volume fraction $\epsilon_i$ of a component is given by
\EQ
\epsilon_i = \frac{\frac{m_i}{\rho_{i}}}{\sum_i \frac{m_i}{\rho_{i}}},
\label{eq:epsilon}
\EN
where $m_i$ is the mass concentration of each component and
$\rho_i$ is the material density of each component.
The mass $m_i$ of major components was measured by
an Aerodyne High Resolution Time-of-Flight Aerosol Mass Spectrometer (HR-ToF-AMS) \citep{decarlo2008fast}.
The uncertainty in measured mass concentrations is up to $50\%$ due to uncertainties in the applied instrument collection efficiency, which was assumed to be unity for this dataset based on preliminary comparison with particle-into-liquid sampler (PILS) measurements. 
We take the time-averaged $m_i$ for each BCB flight leg to calculate
$\epsilon_i$.
The relative mass concentration of organic aerosol, sulfate (${\rm SO}_4^{2-}$),
nitrate (${\rm NO}_3^{-}$), and ammonium (${\rm NH}_4^{+}$) from the AMS measurement are listed in
\Tab{tab:m} and the corresponding mass of ${\rm (NH_4)_2 SO_4}$ and
${\rm NH_4NO_3}$ are listed in \Tab{tab:kappa}.
A mass-weighted $\kappa$ is calculated from AMS-measured organic, sulfate, and nitrate mass and by assuming both sulfate and nitrate are fully neutralized as ${\rm (NH_4)_2 SO_4}$ and
${\rm NH_4NO_3}$ to assign appropriate $\kappa$ values (\Tab{tab:kappa}).  Ammonium concentrations measured by the AMS support this assumption.
Time-averaged $\bar{\kappa}$ calculated according to \Eq{eq:kappa}
is listed in the last column of \Tab{tab:m}.
We use same $\bar{\kappa}$ for different modes of aerosol
size distributions even though \citet{fridlind2017derivation}
indicates that smaller modes have smaller kappas.

\subsection{Droplet/ice size distribution}

Cloud droplet/ice size distribution, liquid or ice water content
(LWC or IWC), number concentration of cloud droplet (ice crystals)
$N_c$ ($N_{\rm ice}$), and effective radius $r_{\rm eff}$ were
measured by Fast Cloud Droplet Probe (FCDP) equipped on
Falcon HU-25. The FCDP measures particles in a size range of $3-50\, \mu{\rm m}$ with an uncertainty of
less than $20\%$ \citep{baumgardner2017cloud, knop2021comparison}. The cutoff is $3.5\, \mu{\rm m}$. The two-dimensional stereo (2DS) probe \citep{lawson20062d} equipped on HU-25 measures size and concentration of cloud/ice particles in the range of $11.4-1464.9\, \mu{\rm m}$ in diameter
with a spatial resolution of $11.4\, \mu{\rm m}$/pixel \citep{voigt2010situ, bansmer2018design}. However, the first size bin of 2DS measurement was excluded due to large uncertainties. Therefore, 2DS measurement covers
a size range of $28.5-1464.9\, \mu{\rm m}$ in the present study.

\subsection{ERA5 and MERRA-2 reanalysis}

As described in Part I \citep{2021arXiv210706193L}, large-scale forcings (i.e.,mositure and temperature advective tendencies and wind profiles) and surface heat fluxes to drive the WRF-LES are obtained from
hourly model-level and surface-level ERA5 reanalysis with a mesh grid-size of 31 km
\citep{hersbach2018era5}. The 3-hourly model-level and 1-hourly surface-level
MERRA-2 \citep{merra-2} reanalysis with a horizontal resolution of $0.5^\circ\times0.625^\circ$ are used to compare with WRF-LES and ERA5.  

\subsection{GOES-16 satellite retrievals}

We use retrievals from the Geostationary Operational Environmental
Satellite (GOES) to evaluate ERA5/MERRA-2 and WRF-LES.
GOES-16 Advanced Baseline Imager (ABI) retrievals (LWP, $N_c$, and $r_{\rm eff}$) are produced
using the Clouds and Earth’s Radiant Energy System (CERES)
Edition 4 algorithms \citep{mi06520w, trepte2019global} adapted
to geostationary satellites \citep{minnis2008near}.
The GOES-16 products have a pixel size of 2 km and a selected time resolution of 20 minutes.

\section{LES numerical experiment setup}
\label{sec:numerical}

The LES setup is the same as in Part I \citep{2021arXiv210706193L}.
The lateral size of the LES domain is $L_x=L_y=60\, \rm{km}$
with a grid spacing of $dx=dy=300\, \rm{m}$. There are 153 vertical layers
up to $z_{\rm top} = 7\, \rm{km}$. The two-moment Morrison cloud
microphysics scheme with prescribed aerosol size modes (see section \ref{sec:obs}.\ref{sec:aerosol}) and hygroscopicity (see section \ref{sec:obs}.\ref{sec:kappa}) is employed
\citep{endo2015racoro}. 
The same time-varying large-scale forcing
(i.e., temperature and water vapor mixing ratio tendencies $\partial_t \bar{\theta}$ and $\partial_t \bar{q}_v$,
wind speed $u \& v$ relaxation, and divergence $\bar{D}$)
and surface heat fluxes described in Part I are adopted.
To investigate the aerosol effects, we perform
simulations with prescribed aerosol size distributions
derived from the ACTIVATE campaign measurements as
described in section~\ref{sec:obs}.\ref{sec:aerosol}. Details of simulations
are listed in \Tab{tab:runs}.

\begin{table*}[t!]
\caption{
List of simulations.
``NC'' denotes prescribed cloud droplet
number concentration and ``NA'' denotes prescribed
aerosol number concentration. BCB flight legs and $\bar{\kappa}$
are consistent with the values listed in \Tab{tab:kappa}.
%Time-varying surface heat fluxes from ERA5 reanalysis
%are adopted for all simulations.
The mean $\kappa$ of each aerosol components ($\bar{\kappa}$) is used
in the simulations unless otherwise specified. 
$\febm$ and $\febu$ denote the simulations with $\bar{\kappa}_{\rm org}$ and $\kappa^{\rm max}_{\rm org}$, respectively.
Simulations $0228\_$NC
and $0301\_$NC are from \citet{2021arXiv210706193L}.
} 
\centering
\setlength{\tabcolsep}{1pt}
\begin{tabular}{|c|c|c|c|c|c|c|}
\hline
Simulations & $N_a$ [$\rm{cm}^{-3}$] & $N_c$ [$\rm{cm}^{-3}$] & BCB leg & $\bar{\kappa}$ \\
\hline
0228$\_$NC & -- & 650 & -- & -- \\ %WRF/lasso_wrf_faster_raw/run_activate_tendencyTQv_rlxUV_7km_Nc_0228_HFt 
$\febm$ & 5593 & -- & BCB1 & 0.313 \\ %WRF/lasso_wrf_faster_raw/run_activate_tendencyTQv_rlxUV_7km_NA_HFt_0228 
$\febu$ & 5593 & -- & BCB1 & 0.392 \\ %WRF/lasso_wrf_faster_raw/run_activate_tendencyTQv_rlxUV_7km_NA_0228_kappa0p2_HFt 
0228$\_{\rm NA}2$ &  5364 & -- & BCB2 & 0.341 \\ %WRF/lasso_wrf_faster_raw/run_activate_tendencyTQv_rlxUV_7km_NA60221_HFt_0228 
0301$\_$NC & -- & 450  & -- & --\\ %WRF/lasso_wrf_faster_raw/run_activate_tendencyTQv_rlxUV_7km_Nc_HFt 
0301$\_$NA1 & 1434 & -- & BCB1 & 0.451 \\ %WRF/lasso_wrf_faster_raw/run_activate_tendencyTQv_rlxUV_7km_NA_HFt
0301$\_{\rm NA2}$  &  3100 & -- & BCB2 & 0.479 \\ %WRF/lasso_wrf_faster_raw/run_activate_tendencyTQv_rlxUV_7km_NA57081_HFt
\hline
\multicolumn{5}{p{0.3\textwidth}}{}
\end{tabular}
\label{tab:runs}
\end{table*}

\section{Aerosol-meteorology-cloud interactions}
\label{sec:LES}

The ultimate goal of this study is to investigate how aerosols
affect cloud micro-/macro-physical processes under different
meteorological conditions of the two CAO cases.
We first compare the cloud microphysical properties between WRF-LES
and FCDP measurements. Then, aerosol effects on LWC, CFC, and radiation
are investigated.
Finally, we address how aerosols impact the BL meteorology. 

\subsection{Comparison of cloud microphysical properties}

\begin{figure*}[t!]\begin{center}
\begin{overpic}[width=\textwidth]{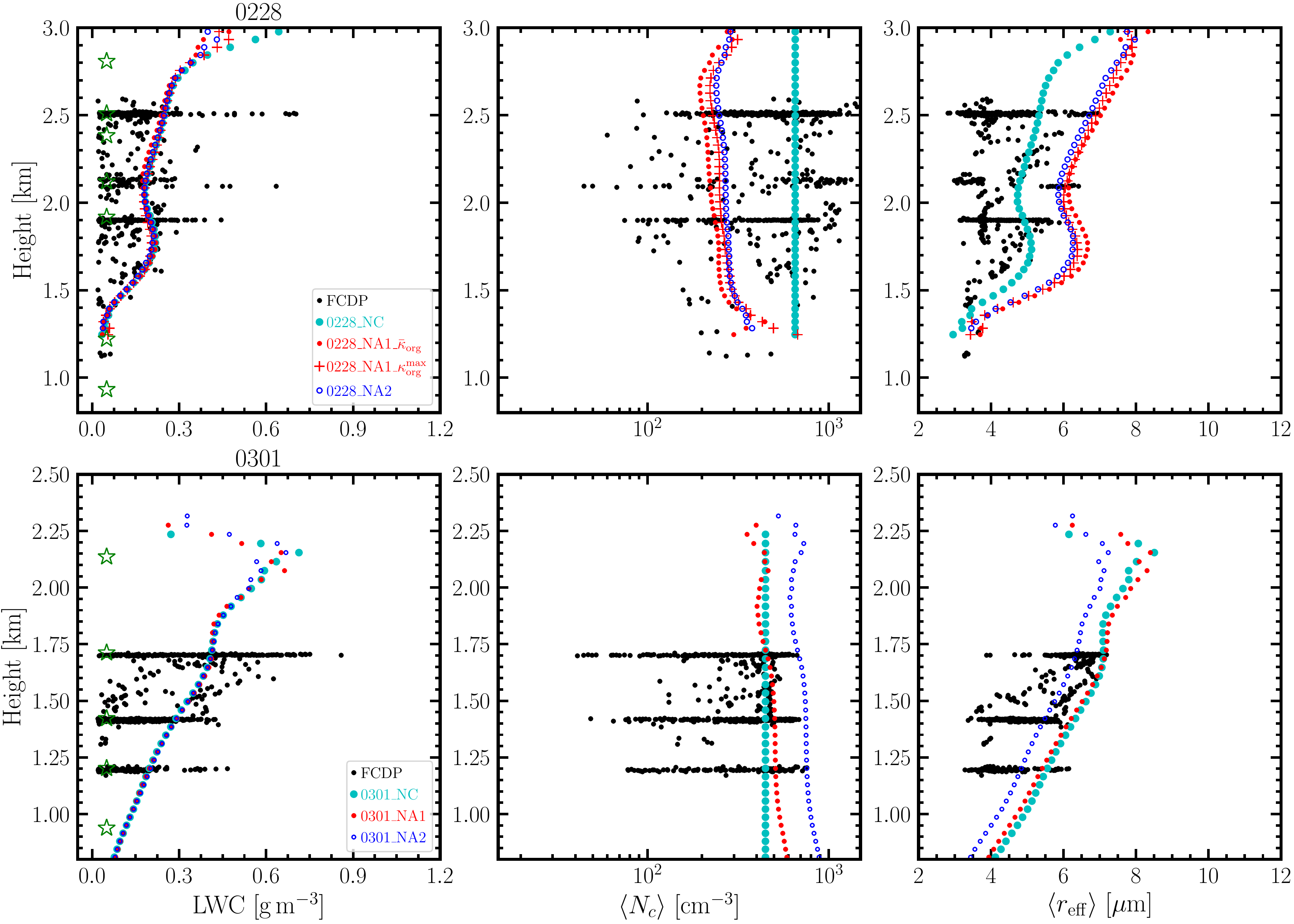}\put(-1,68.5){(a)}\put(-1,34){(b)}\end{overpic}
\end{center}\caption{Comparison of vertical profiles of LWC, $\langle N_c
\rangle$, and $\langle r_{\rm eff} \rangle$ between the WRF-LES
and the FCDP sampling.
A threshold of ${\rm LWC}=0.02\, \rm{g\, m}^{-3}$, $3.5\,\mu{\rm m} \le d_{\rm eff}
\le 50\,\mu{\rm m}$, and $N_c=20\, \rm{cm}^{-3}$ is applied to both
the WRF-LES and the FCDP sampling (black dots).
For the WRF-LES,
only grid cells within clouds (vertically and laterally) are averaged
to obtain the vertical profile.
The measurement took place between 16:00 UTC to 17:00 UTC
for the February 28 and 15:00 UTC to 16:00 UTC for the March 1 cases.
The corresponding mean vertical profile of LWC, $\langle N_c
\rangle$, and $\langle r_{\rm eff} \rangle$ 
is obtained by averaging three snapshots of WRF-LES output as the output
frequency is 30 minutes.
The green stars mark all the above cloud base (ACB) and below cloud top (BCT) flight legs.
}
\label{FCDP_WRF_comp}
\end{figure*}

\begin{figure*}[t!]\begin{center}
\begin{overpic}[width=\textwidth]{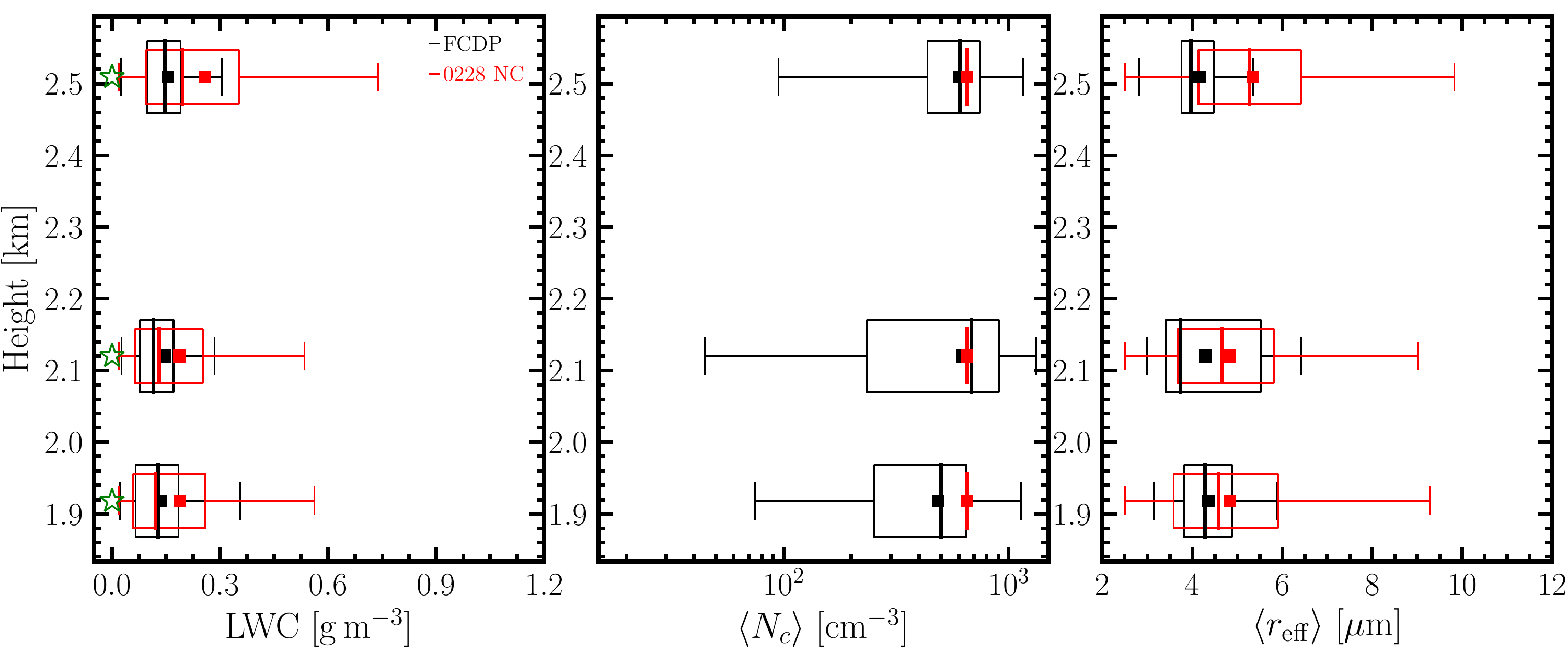}\put(0,40){(a)}\end{overpic}
\begin{overpic}[width=\textwidth]{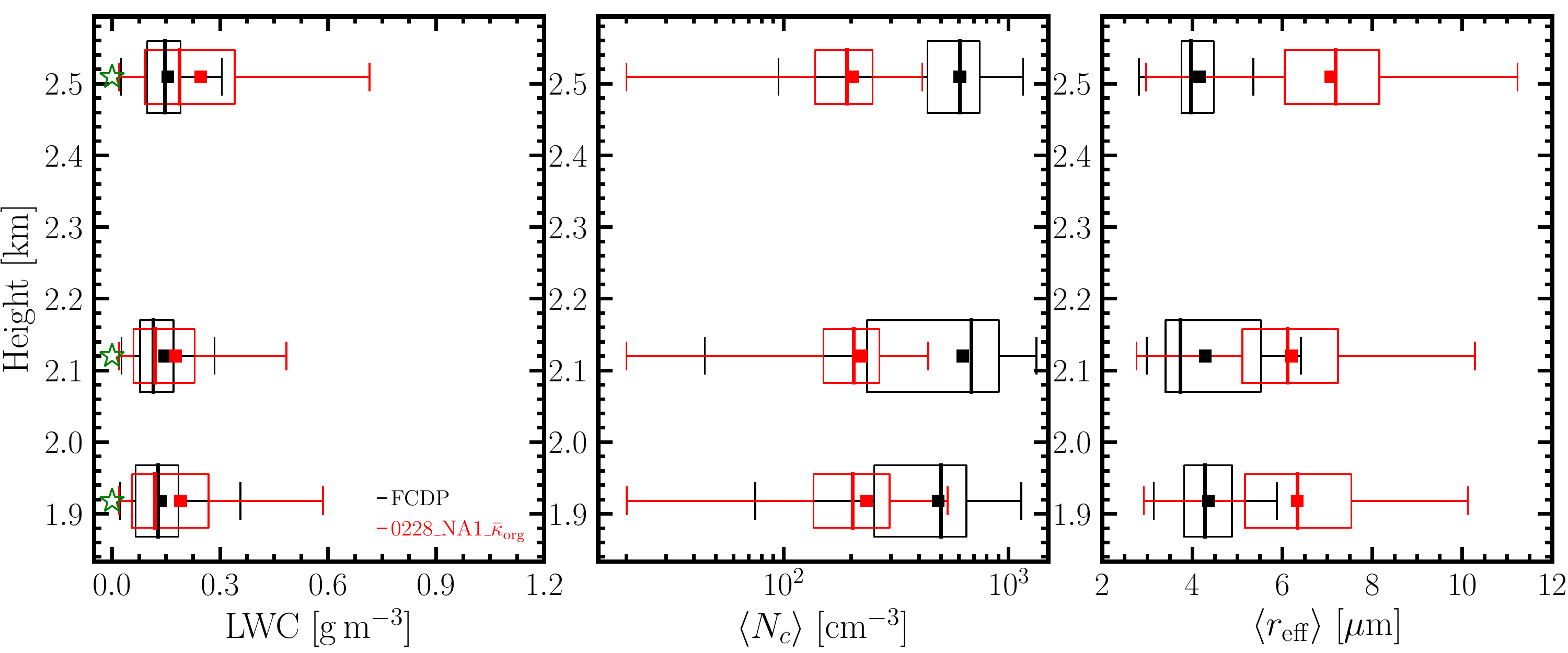}\put(0,40){(b)}\end{overpic}
\end{center}\caption{Corresponding statistics of \Fig{FCDP_WRF_comp}(a) for the
February 28 case with (a) constant $N_c$ (simulation 0228$\_$NC) and (b)
constant $N_a$ (simulation $\febm$).  Only flight legs (all ACB and BCT
marked by green stars)  within clouds that have sufficient data are used.  The
data are binned at those heights with a residual range of $\pm 50\, {\rm m}$ such
that at least one model layer is counted at the height of each flight legs.
Smaller residual ranges does not affect the statistics.  The height of the red
box represents the bin width and the one of the black box is rescaled for
readability.  In the box-and-whisker plot, the binned data extends horizontally
from the 25th (Q1, l.h.s wall of the box) to the 75th (Q3, r.h.s wall of the
box) percentile with the median represented by the splitting line inside the
box, the mean represented by solid squares inside the box, the minimum ($Q_{\rm
min}$) and maximum ($Q_{\rm max}$) values represented by the left and right end
of whiskers, respectively.  Here $Q$ denotes values of a quantity (i.e., LWC,
$\langle N_c \rangle$, and $\langle r_{\rm eff} \rangle$).  The statistics of
$N_c$ are a single point as $N_c$ is constant in (a). 
}
\label{FCDP_0228}
\end{figure*}

We start with
control simulations with constant cloud droplet number concentration $N_c$ ($0228\_$NC
and $0301\_$NC in \Tab{tab:runs}), which
are compared to simulations adopting prescribed aerosol
size distributions derived from in-situ measurements during different BCB flight legs. The constant $N_c$
is estimated from the FCDP measurements during the in-cloud legs of February 28 and March 1 cases.
%The WRF-LES results are then compared to the FCDP sampling.
\Fig{FCDP_WRF_comp} shows the comparison between WRF-LES
and FCDP vertical profiles of LWC,
$\langle N_c\rangle$, and $\langle r_{\rm eff} \rangle$.
For the February 28 case, simulation 0228$\_$NC and $\febm$
yields almost the same LWC and different $\langle N_c\rangle$
and $\langle r_{\rm eff} \rangle$ between 16:00-17:00 UTC.
By examining the corresponding statistics shown in
\Fig{FCDP_0228}(a) at flight legs
with sufficient statistics, it is evident that
simulation 0228$\_$NC is able to capture the vertical structure of
$\langle N_c\rangle$ and $\langle r_{\rm eff} \rangle$ while
$\febm$ underestimates them compared to the FCDP measurements.

The hygroscopiticy of organic aerosols $\kappa_{\rm org}$ is difficult to determine due to uncertainties of their sources. 
We perform two simulations ($\febm$ and $\febu$)
with the same prescribed aerosol size distributions but different
$\bar{\kappa}$ (due to different $\kappa_{\rm org}$) estimated from the in-situ aircraft measurements of mass of aerosol components during the ACTIVATE campaign for the February 28 case. As discussed in
section~\ref{sec:obs}.\ref{sec:kappa}, we derive $\bar{\kappa}$
from either the mean or largest value of $\kappa_{\rm org}$.
$\kappa_{\rm org}=0.1$, a mean value from Table 1 of
\citet{petters2007single}, is used to calculate $\bar{\kappa}$ for simulation $\febm$.
In simulation $\febu$, the upper value $\kappa_{\rm org}=0.229$ is adopted.
As shown in \Fig{FCDP_WRF_comp}, $\langle N_c\rangle$ and
$\langle r_{\rm eff} \rangle$ from simulation $\febu$ are almost identical
to the ones from $\febm$.
This is because $\kappa$ of the organic component is about six times small than of the non-organic components even
though the mass fraction of the organic component
is $54.5\%$. 

We then investigate how vertical profiles of
$\langle N_c\rangle$ and $\langle r_{\rm eff} \rangle$ depend on
different aerosol size distributions obtained from two BCB flight legs
for the February 28 case. Simulation $0228\_\rm{NA}2$ is the same as
$\febm$ but with aerosol size distributions (\Fig{dNdlnD_53602_54105_2modes_0301}(b)) and $\bar{\kappa}$
derived from BCB2 (\Tab{tab:fit} and \Tab{tab:kappa}).
\Fig{FCDP_WRF_comp}(a) shows that
$\langle N_c\rangle$ and $\langle r_{\rm eff} \rangle$ from
simulation $\febm$ (red dots) are very close to the ones
from $0228\_\rm{NA}2$ (blue open circles), which is due to the fact
that aerosol size distributions from the two BCB flight legs
are similar as shown in \Fig{dNdlnD_60221_60423_3modes_0228}.
It is worth noting that the aerosol (LAS and SMPS) and FCDP measurements
were not carried out simultaneously or collocated  as shown in
\Fig{cpc_0228}(a) for neither BCB flight legs.
\Fig{FCDP_traj}(a) shows the Falcon flight trajectory with the contour
level representing the measurement time and stars and squares marking
the start and end of BCB1 and BCB2 for the February 28 case.
Most of the sampling of FCDP took place around 16:24 UTC (\Fig{cpc_0228}(a)), which
is closer to the BCB2 flight leg (\Fig{FCDP_traj}(a)).  
This suggests that the location of aerosol measurements
during BCB2 is closer to the FCDP measurement than
that during BCB1, which explains
why $\langle N_c\rangle$ and $\langle r_{\rm eff} \rangle$ from
simulation $0228\_\rm{NA}2$ are slightly closer to FCDP measurements than
those from $\febm$.
Therefore, the discrepancy of $\langle N_c\rangle$ and
$\langle r_{\rm eff} \rangle$ between ``NA'' simulations and
FCDP measurements likely depend on the collocation between the
aerosol measurements and the FCDP measurements.

\begin{figure*}[t!]\begin{center}
\begin{overpic}[width=\textwidth]{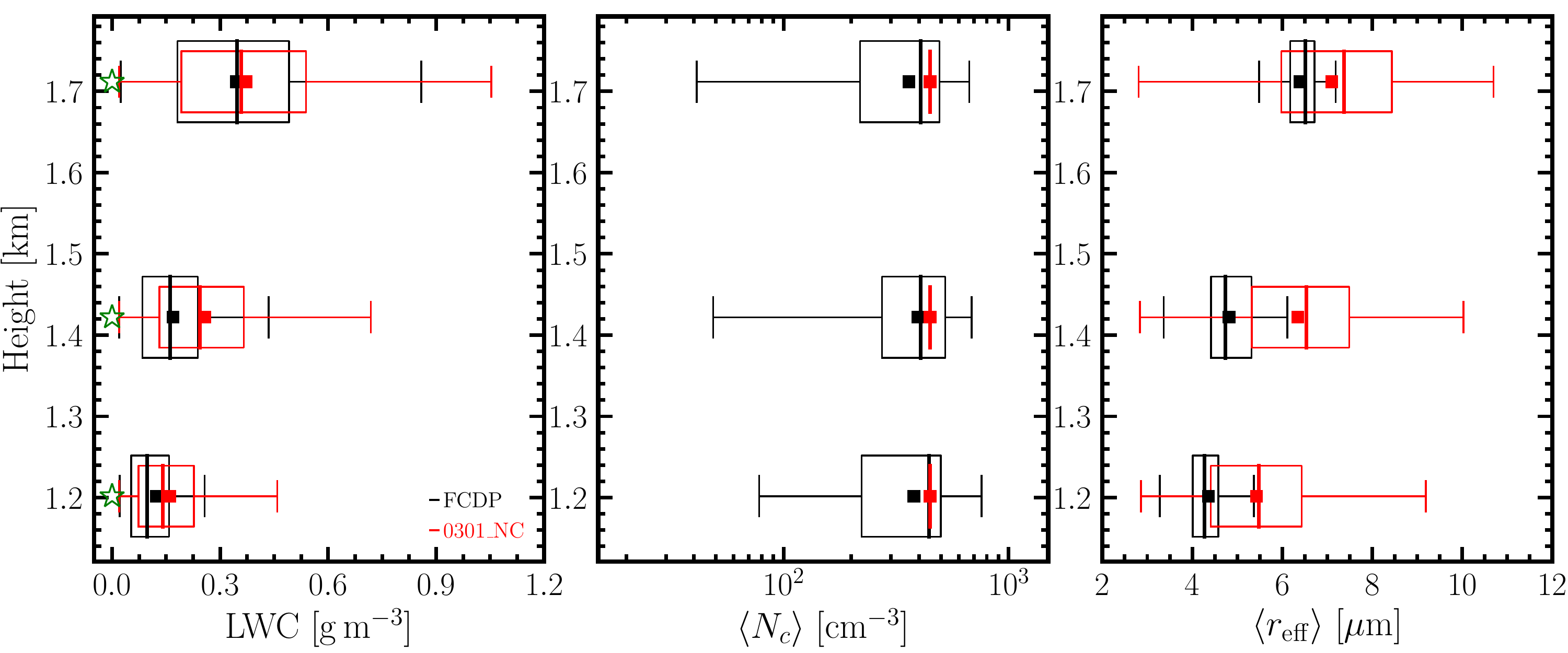}\put(0,40){(a)}\end{overpic}
\begin{overpic}[width=\textwidth]{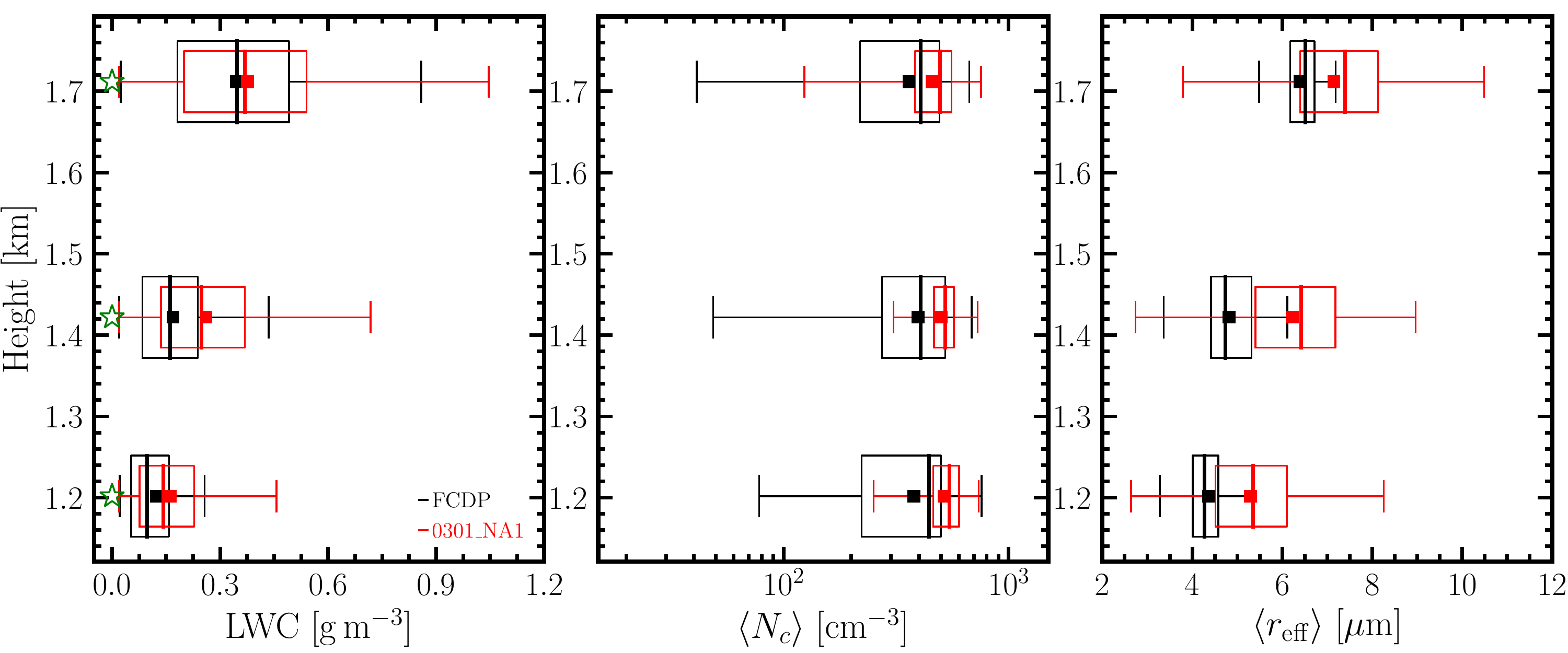}\put(0,40){(b)}\end{overpic}
\end{center}\caption{Same as \Fig{FCDP_0228} but for the March 1 case:
simulation (a) $0301\_\rm{NC}$ and (b) $0301\_\rm{NA1}$.
}
\label{FCDP_0228_NA}
\end{figure*}

\Fig{FCDP_WRF_comp}(b) shows the comparison for the
March 1 case. Vertical profiles from simulation $0301\_\rm{NC}$
are close to the ones from $0301\_\rm{NA}1$.
The statistics of vertical profiles from both simulation $0301\_\rm{NC}$
and $0301\_\rm{NA1}$ agree well with the FCDP measurements
as shown in \Fig{FCDP_0228_NA}. We also perform a simulation with
aerosol size distributions from BCB2 shown in \Fig{dNdlnD_53602_54105_2modes_0301}(b), the integrated aerosol number concentration
($\bar{N}_{\rm a}$) of which is about two times
larger than the one from aerosol size distributions shown in \Fig{dNdlnD_53602_54105_2modes_0301}(a). As expected,
$\langle N_c\rangle$ from simulation $0301\_\rm{NA}2$
is larger than the one from $0301\_\rm{NA}1$ as shown in
\Fig{FCDP_WRF_comp}(b). This is because the BCB1 flight leg
is co-located right below the FCDP sampling while the BCB2 flight leg
is not collocated well with the FCDP sampling as shown in \Fig{FCDP_traj}(b) (the contour line between stars and the thick magenta line).

We also examine the diurnal variation of cloud properties by
comparing the LES with the FCDP measurements during flights from 20:47:07 to 21:10:58 UTC and from 19:43:30 to 20:13:44 UTC
for the February 28 and March 1 cases, respectively.
LWC from the LES and FCDP measurement agree well with each other
(\Fig{FCDP_WRF_comp_L2}(a) and \Fig{FCDP_L2_0228})
for the February 28 case. $N_c$ from the NA simulations capture the
FCDP measurement better than the NC simulation. Interestingly,
$r_{\rm eff}$ from the FCDP measurement is captured well by
the NC simulation between $1.7-2.5\, \rm{km}$ and is closer to NA simulations
above $2.5\, \rm{km}$. 
For the March 1 case, the LES capture the FCDP
measurements well except that it slightly
underestimates $r_{\rm eff}$ (\Fig{FCDP_WRF_comp_L2}(b) and \Fig{FCDP_L2_0301}). 
Overall, the LES is able to capture the diurnal
variation of cloud properties.

\subsection{Impact of different aerosol treatments on clouds and radiation}

\begin{figure*}[t!]\begin{center}
\includegraphics[width=\textwidth]{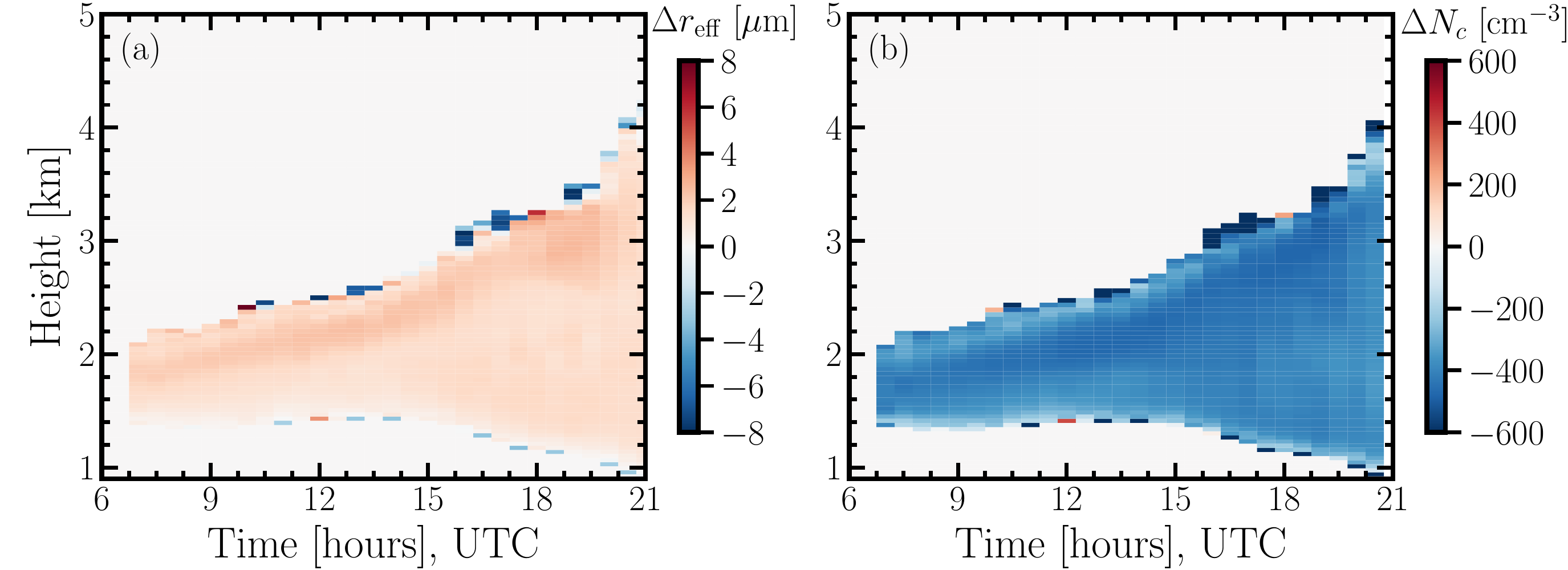}
\end{center}\caption{Evolution of difference of
(a) $r_{\rm eff}$
and (b) $N_c$ profiles between
simulation $\febm$ and 0228$\_$NC averaged during the dropsonde
measurement time 16:00-17:00 UTC for the February 28 case. 
}
\label{re_t_diff_0228}
\end{figure*}

\begin{figure*}[t!]\begin{center}
\includegraphics[width=\textwidth]{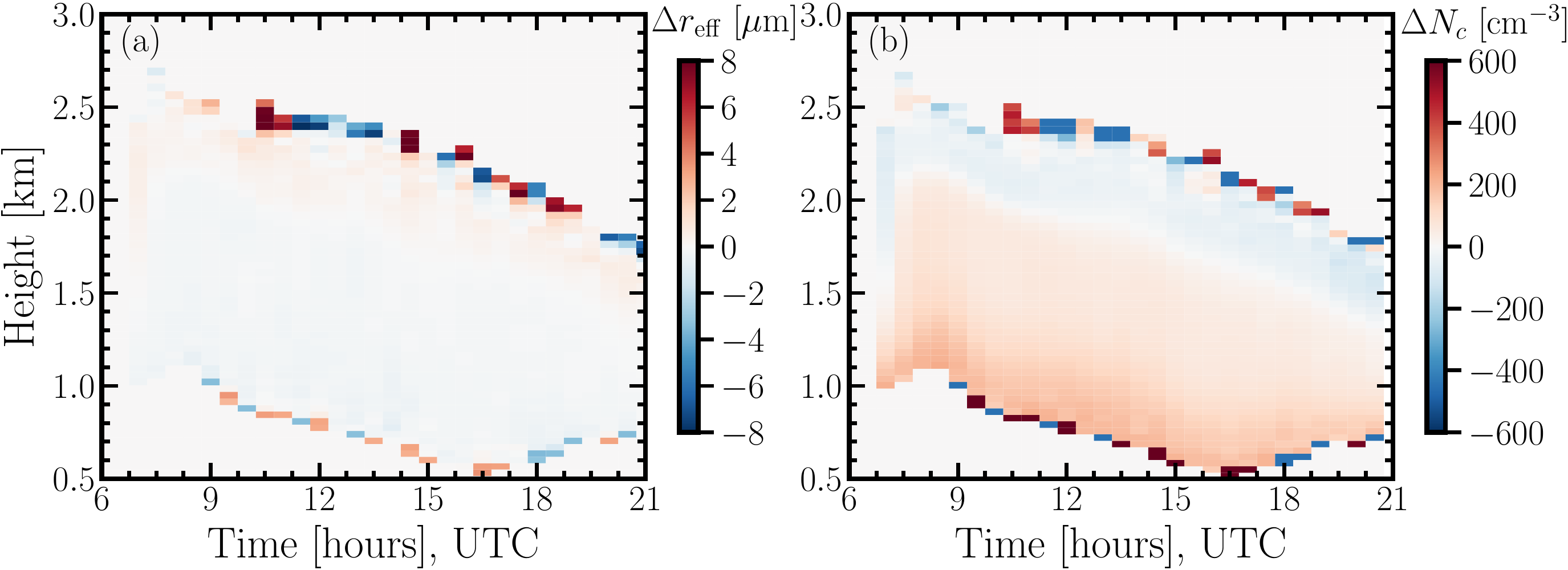}
\end{center}\caption{Same as \Fig{re_t_diff_0228} but for the March 1 case (simulation 
0301$\_$NA1 and 0301$\_$NC averaged between 15:00 and 16:00 UTC). 
}
\label{re_t_diff_0301}
\end{figure*}

The comparison to in-situ cloud observations indicates that the WRF-LES simulations can reasonably capture the
vertical distribution of LWC in both CAO cases. Within the aerosol measurement uncertainties, the prescribed aerosol size distributions to WRF-LES show different impacts on $N_c$ and other cloud properties between February 28 and March 1.
In both cases, the LWC shows very small sensitivity
to different aerosol size distributions. Therefore, changes in $r_{\rm eff}$
and $N_c$
primarily reflect the impact of aerosols on droplet nucleation, similar to the first indirect effect. \Fig{re_t_diff_0228} shows the difference of vertical profiles
of $r_{\rm eff}$ and $N_c$ between simulation $\febm$ and $0228\_{\rm NC}$.
Interestingly, both $\Delta r_{\rm eff}$ and $\Delta N_c$ exhibit
monotonic patterns vertically with a relatively larger sensitivity near cloud top.
\Fig{re_t_diff_0301} shows $\Delta r_{\rm eff}$ and $\Delta N_c$ for the March 1 case.
Except for the large inhomogeneity at the cloud top and base, there is a different cloud droplet (number and size) response to the prescribed aerosols near cloud base.

Ice particles were observed for the March 1 case but were barely detected for
the February 28 case as shown in \Fig{2DS_WRF_ice}, where vertical profiles of
IWC and $\langle N_{\rm ice}\rangle$ from WRF-LES and 2DS measurements are
shown, respectively.  $r_{\rm eff, ice}$ from 2DS sampling is also shown.
Neither IWC nor $\langle N_{\rm ice}\rangle$ are sensitive to aerosols in this
case study. \Fig{2DS_0301_NA} shows the corresponding statistics of IWC and
$\langle N_{\rm ice}\rangle$. $\langle N_{\rm ice}\rangle$ from the WRF-LES
agrees reasonably well with the one from the 2DS sampling while IWC is off;
however, the WRF-LES captures the higher magnitude of IWC in March 1. 

\begin{figure*}[t!]\begin{center}
\includegraphics[width=\textwidth]{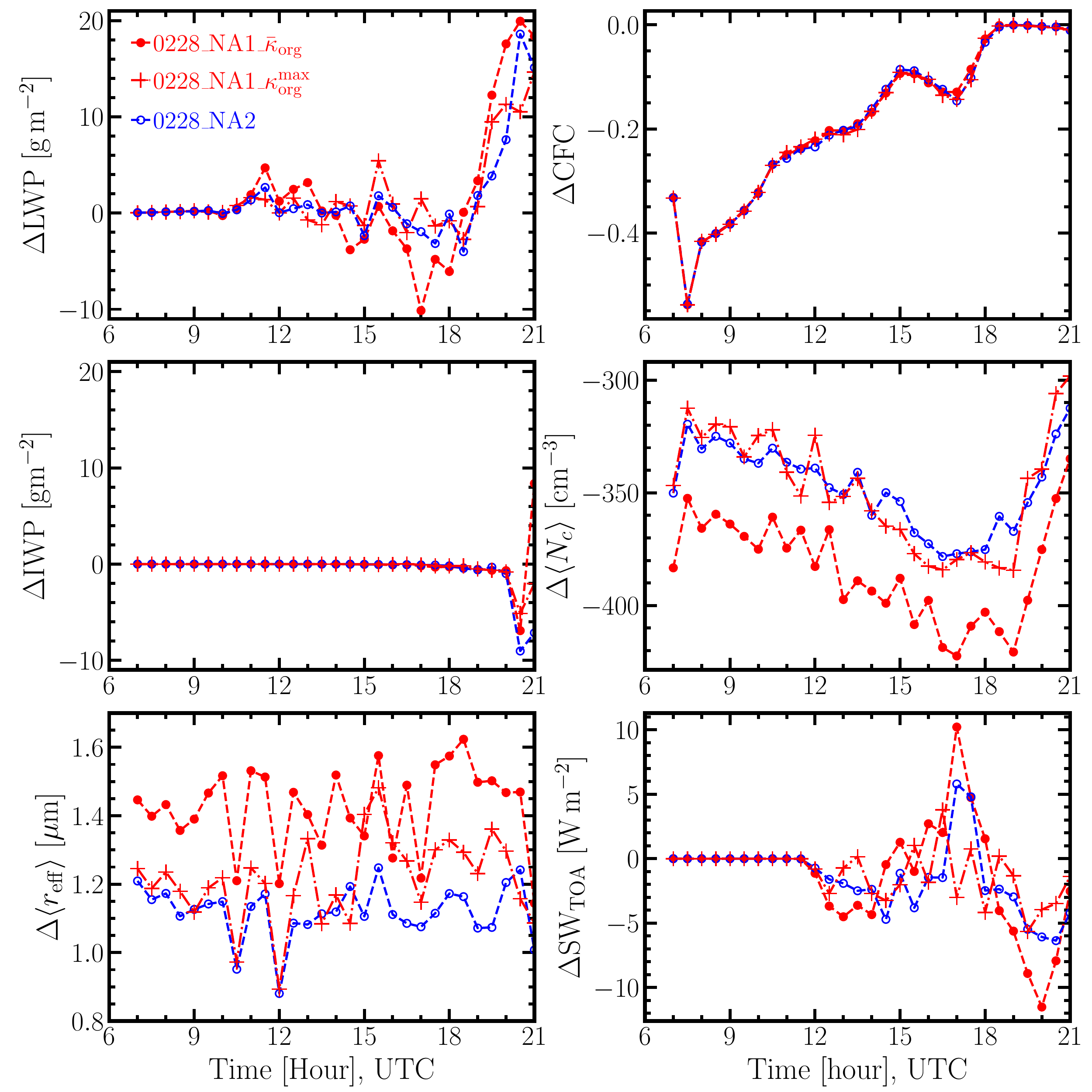}
\end{center}\caption{Time-series differences of LWP, CFC, IWP,
$\langle N_c \rangle$, $\langle r_{\rm eff} \rangle$,
and $\rm{SW_{TOA}}$ between the control simulation (0228$\_$NC)
and the ones ($\febm$, solid red circles; $\febu$, red pluses; 0228$\_$NA2, open blue circles) with prescribed
aerosol size distributions for the February 28 case.
$\langle N_c\rangle$ and $\langle r_{\rm eff}\rangle$ are averaged within clouds ($q_c\ge q_c^*$).
The threshold $q_c^{\star}=0.02 \rm{g\, kg^{-1}}$ is adopted to define the cloud fractional coverage (CFC).
CFC is calculated by counting the vertical columns where $q_c \ge q_c^{\star}$, which is then
normalized by the number of total vertical column of the entire domain.
LWP includes liquid water and rain. IWP includes ice, graupel, and snow.
}
\label{wp_comp_0228}
\end{figure*}

To further quantitatively examine the sensitivity of cloud macro-physical properties
(LWP, CFC, and IWP) and micro-physical properties ($N_c$ and $r_{\rm eff}$)
to the prescribed aerosols, we compare results between simulations with prescribed $N_c$ and aerosol size distributions.
To quantify the impact of prescribed aerosols
in both the sign and the magnitude, we use the metric of percentage difference (PD),
defined as ${\rm PD}=(\mathcal{Q}_{\rm NA}-\mathcal{Q}_{\rm NC})/\mathcal{Q}_{\rm NC}\times 100\%$ with
$\mathcal{Q}_{\rm NC}$ and $\mathcal{Q}_{\rm NA}$ representing quantities from the NC (baseline)
and NA simulations, respectively. $\mathcal{Q}$ is averaged
between 12:00 and 18:00 UTC.
For the February 28 case, we also assess the sensitivity of cloud and radiation to $\kappa_{\rm org}$.
The time evolution of differences in liquid and ice water path, cloud fraction, $\langle N_c\rangle$,
$\langle r_{\rm eff}\rangle$, and 
short-wave (SW) cloud forcing at the top of atmosphere
($\rm{SW_{TOA}}$) between the control simulation 0228$\_$NC
and the ones with prescribed aerosol size distributions
($\febm$, $\febu$, and 0228$\_$NA2) (based on different BCB flight legs)
for the February 28 case is shown in \Fig{wp_comp_0228}.
We first examine how much NA simulations differ from the 
NC simulation. LWP from the NA simulations only changes slightly
compared to the NC simulation (PD=$-2.6\%$, $0.6 \%$, and $-0.5\%$ for simulations $\febm$, $\febu$, and $022\_{\rm NA}2$, respectively) while IWP decreases considerably (PD=$-24.8\%$, $-11.6 \%$, and $-12.4\%$). 
The CFC also decreases (PD=$-14.7\%$, $-15.1 \%$, and $-14.9\%$).
$N_c$ decreases substantially (PD=$-60.9\%$, $-55.7 \%$, and $-55.0\%$).
Correspondingly, $\langle r_{\rm eff} \rangle$ increases (PD=$27.2\%$, $23.8 \%$, and $21.5\%$).
$\Delta \rm{SW_{TOA}}$ changes by $0.19\, \rm{W\, m}^{-2}$, $-0.94\, \rm{W\, m}^{-2}$,
and $-0.93\, \rm{W\, m}^{-2}$
for simulations $\febm$, $\febu$, and $022\_{\rm NA}2$, respectively.

To examine the effect of $\kappa$ on these quantities,
we compare simulation $\febm$ and $\febu$.
We use $\Delta {\rm PD}=(\mathcal{Q}_{\rm NA_i}-\mathcal{Q}_{\rm NA_j})/\mathcal{Q}_{\rm NC}$ to quantify differences between NA simulations,
where the subscripts $\rm NA_i$ and $\rm NA_j$ indicate two different
NA simulations.
$\Delta {\rm LWP}$ from simulations $\febm$ (red solid circles) and $\febu$ (red pluses) starts to differ around 10:00 UTC while $\Delta {\rm IWP}$
starts to differ around 15:00 UTC (\Fig{wp_comp_0228}). The difference of PD between simulation
$\febu$ and $\febm$ in LWP is $\Delta \rm{PD_{LWP}}=(0.63-(-2.58))\%=3.21\%$,
and in IWP is $\Delta \rm{PD_{IWP}}=13.11\%$. 
However, $\Delta \rm{CFC}$ is almost the same with $\Delta \rm{PD_{CFC}}=-0.44\%$. $\Delta \langle N_c \rangle$
and $\Delta \langle r_{\rm eff} \rangle$ shows considerable difference between
simulations $\febm$ and $\febu$ with $\Delta \rm{PD_{N_c}}=5.22\%$
and $\Delta \rm{PD_{r_{eff}}}=-3.41\%$, respectively.
$\Delta \rm{SW_{TOA}}$ decreases by
$1.13\, \rm{W\, m}^{-2}$.

How cloud properties respond to aerosol size distributions
from different BCB flight legs is examined by comparing simulations $\febm$ (red solid circles) and
0228$\_$NA2 (blue open circles) in \Fig{wp_comp_0228}, in which aerosol size distributions from BCB2 listed in
\Tab{tab:fit} is used. Minor differences are observed for $\Delta {\rm LWP}$
($\Delta \rm{PD_{LWP}}=1.06\%$)
while considerable differences are evident
for $\Delta {\rm IWP}$ ($\Delta \rm{PD_{IWP}}=12.37\%$), $\langle N_c \rangle$ ($\Delta \rm{PD_{N_c}}=5.85\%$)
and $\langle r_{\rm eff} \rangle$ ($\Delta \rm{PD_{r_{eff}}}=-5.74\%$).
$\Delta \rm{CFC}$ ($\Delta \rm{PD_{CFC}}=-0.16\%$)
almost has no difference. However, the magnitude of the CFC vertical-structure exhibits difference as shown in \Fig{cfc0228}. 
$\Delta \rm{SW_{TOA}}$ decreases by $1.11\, \rm{W\, m}^{-2}$. We further quantify the aerosol effect on radiative forcing by examining
the response of cloud optical depth $\tau_{\rm c}$ to cloud-top $N_c$ via the
following relation \citep{ghan2016challenges},
\EQ
\label{eq:tau}
\frac{\Delta \ln \overline{\tau_{\rm c}}}{\Delta \ln \overline{\langle N_c\rangle}}=\frac{\Delta \ln \overline{\rm LWP}}{\Delta \ln \overline{\langle N_c\rangle}}-\frac{\Delta \ln \overline{\langle r_{\rm eff}\rangle}}{\Delta \ln \overline{\langle N_c\rangle}}.
\EN
Perturbations of LWP and cloud-top $r_{\rm eff}$ due to $N_c$ (cloud-top) are $\Delta \ln \overline{\rm LWP}/\Delta \ln \overline{\langle N_c\rangle}=0.150$
and $\Delta \ln \overline{\langle r_{\rm eff}\rangle}/\Delta \ln \overline{\langle N_c\rangle}=-0.269$, respectively, which leads to $\Delta \ln \overline{\tau_{\rm c}}/\Delta \ln \overline{\langle N_c\rangle}=0.419$ according to \Eq{eq:tau}.

For the March 1 case,
LWP only changes slightly with PD=$0.06\%$ and $-0.02\%$ for simulations $0301\_{\rm NA1}$
and $0301\_{\rm NA2}$ compared to simulation $0301\_{\rm NC}$, respectively.
PD of IWP is $-0.06\%$ and $2.34\%$.
The CFC decreases with PD=$-3.40 \%$ and $-3.36\%$. 
The magnitude of the CFC vertical-structure is quite
similar as shown in \Fig{cfc0301}.
$\langle N_c\rangle$ increases with PD=$12.80 \%$ and $70.06\%$.
$\langle r_{\rm eff} \rangle$ decreases with PD=$-1.77 \%$ and $-12.77\%$.
$\Delta \rm{SW_{TOA}}$ decreases by $1.24\, \rm{W\, m}^{-2}$ and $0.51\, \rm{W\, m}^{-2}$.
%with PD=$-0.46 \%$ and $-0.19\%$.

\begin{figure*}[t!]\begin{center}
\includegraphics[width=\textwidth]{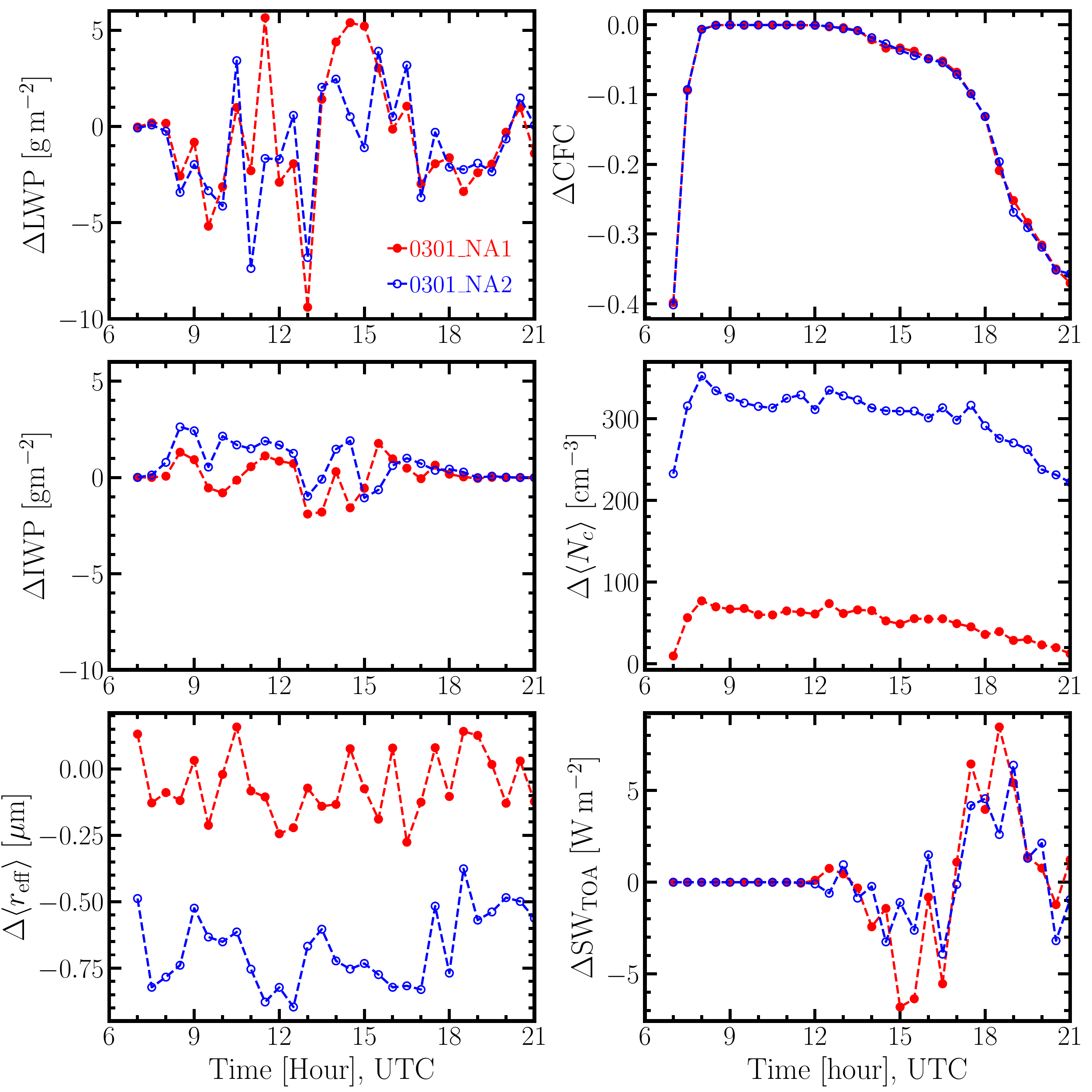}
\end{center}\caption{Same as \Fig{wp_comp_0228} but for the March 1 case.
}
\label{wp_comp_0301}
\end{figure*}

Similar to the February 28 case, 
$\Delta {\rm LWP}$ ($\Delta {\rm PD_{LWP}}=-0.51\%$ between
simulations $0301\_{\rm NA2}$ and $0301\_{\rm NA1}$) and 
$\Delta {\rm IWP}$ ($\Delta {\rm PD_{IWP}}=2.38\%$) in March 1 are slightly affected by different prescribed
aerosol size distributions while CFC ($\Delta {\rm PD_{CFC}}=-0.15\%$) is insensitive to them.
Aerosol size distributions 
are quite different (\Fig{dNdlnD_53602_54105_2modes_0301})
between the two BCB legs
(total $N_a$ from the flight leg BCB2 is about 2.16 times larger
than that of BCB1) for the March 1 case,
which leads to substantial differences in $\langle N_c\rangle$
($\Delta {\rm PD_{N_c}}=57.26\%$) and
$\langle r_{\rm eff}\rangle$ ($\Delta {\rm PD_{r_{eff}}}=-11.00\%$)
as shown in \Fig{wp_comp_0301}.
However, the difference between simulation $0301\_{\rm NA2}$
and $0301\_{\rm NA1}$ in $\Delta \rm{SW_{TOA}}$ is only
$0.72\, \rm{W\, m}^{-2}$. Differences in aerosol size distribution
induced susceptibility of LWP and cloud-top $r_{\rm eff}$ to $N_c$ (cloud-top) are $\Delta \ln \overline{\rm LWP}/\Delta \ln \overline{\langle N_c\rangle}=-0.002$
and $\Delta \ln \overline{\langle r_{\rm eff}\rangle}/\Delta \ln \overline{\langle N_c\rangle}=-0.318$, respectively, which results in a positive perturbation of
$\Delta \ln \overline{\tau_{\rm c}}/\Delta \ln \overline{N_c}$ (\Eq{eq:tau}) of 0.316. 
The susceptibility of LWP and cloud-top $r_{\rm eff}$ to $N_c$ (cloud-top) for both cases are summarized in \Tab{tab:sus}.
The February 28 case yields $\Delta \ln \overline{\rm LWP}/\Delta \ln
\overline{\langle N_c\rangle}=0.15$, which is close to 0.11 reported in
\citet{lee2009aerosol} and summarized in Table S1 of
\citet{glassmeier2021aerosol}.
Our values are very different from other LES studies reported in Table S1 of \citet{glassmeier2021aerosol}, which summarized $\Delta \ln \overline{\rm LWP}/\Delta \ln \overline{\langle N_c\rangle}$ due to precipitation or
entrainment.  
However, we note that both the February 28 and March 1 cases represent non-precipitating
stratocumulous clouds. It is also unclear whether $\Delta \ln \overline{\rm LWP}/\Delta \ln \overline{\langle N_c\rangle}$
is due to entrainment in these two cases. 
Overall, these two case-studies suggest that spatial-temporal variation of aerosol distributions
have a profound effect on $\langle N_c \rangle$ and $\langle r_{\rm eff} \rangle$. The effects on $\Delta {\rm LWP}$, $\Delta {\rm IWP}$,
and $\Delta \rm{CFC}$ are less obvious. 

\begin{table*}[t!]
\caption{Aerosol size distribution
induced susceptibility of LWP and cloud-top $r_{\rm eff}$ to $N_c$ (cloud-top) for the February 28 (0228$\_{\rm NA}2$- $\febm$) and March 1 (0301$\_{\rm NA2}$- 0301$\_$NA1) cases,
respectively.
} 
\centering
\setlength{\tabcolsep}{1pt}
\begin{tabular}{|c|c|c|c|c|c|}
\hline
Case & $\Delta \ln \overline{\rm LWP}/\Delta \ln \overline{\langle N_c\rangle}$ & $-\Delta \ln \overline{\langle r_{\rm eff}\rangle}/\Delta \ln \overline{\langle N_c\rangle}$ & $\Delta \ln \overline{\tau_{\rm c}}/\Delta \ln \overline{\langle N_c\rangle}$ \\
\hline
February 28 & 0.150   & 0.269 & 0.419 \\  
March 1     &  -0.002 & 0.318 & 0.316 \\ 
\hline
\multicolumn{4}{p{0.3\textwidth}}{}
\end{tabular}
\label{tab:sus}
\end{table*}

\subsection{Impact of ACI on the boundary layer meteorology}

The February 28 case is characterized by a warmer, drier and
deeper boundary layer than the March 1 case \citep{2021arXiv210706193L} as shown in \Fig{verticalP_comp}.
It also shows that LES is able to capture the diurnal variation of the meteorological
state (by comparing the LES to dropsonde measurements during the morning
and afternoon flights as shown in \Fig{verticalP_comp}) and cloud properties
for both cases.
In this section, we examine how aerosols impact  
BL meteorology for these two contrasting CAO cases.  

\begin{figure*}[t!]\begin{center}
\includegraphics[width=\textwidth]{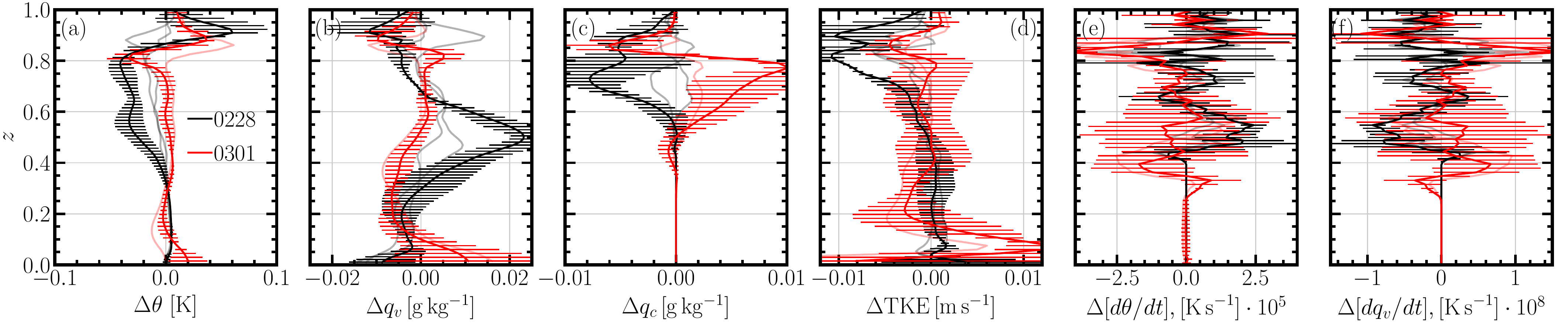}
\end{center}\caption{Difference of vertical profiles between
simulation $\febm$ and 0228$\_$NC (black curves) and between 
0301$\_$NA1 and 0301$\_$NC (red curves) averaged during the dropsonde
measurement time (16:00-17:00 UTC for the February 28 case and
15:00-16:00 UTC for the March 1 case). The error bars indicate $\pm\sigma$ deviation from the time-averaged vertical profiles. The gray and orange curves represent
other simulations.
$z$ is the normalized height by the height of the cloud top.
$\Delta d\theta/ dt$ and $\Delta d q_v/dt$ are due to condensation/evaporation.
Data is only visualized until the cloud top. The vertical profiles
are shown in \Fig{verticalP_comp}.
}
\label{vp_diff}
\end{figure*}

Differences of vertical profiles averaged within the corresponding dropsonde circle
between NA and NC simulations are shown in \Fig{vp_diff}, where $z$ is
the altitude normalized by the cloud top height.
We take simulations $\febm$ and 0301$\_$NA1 as an example.
For both cases, simulations with prescribed aerosol size distributions
yields larger $\theta$ (red and black curves in \Fig{vp_diff}(a)) and smaller $q_v$  (\Fig{vp_diff}(b)) near the inversion top
due to the enhanced $d\theta/ dt$ (\Fig{vp_diff}(e)) and weakened $d q_v/ dt$ (\Fig{vp_diff}(f)) by condensation/evaporation,  but there are quite different responses near cloud base between the two cases.
This demonstrates that aerosol affects meteorological
states via the case-dependent response in cloud condensation/evaporation process to aerosol perturbations.
Even for the same case (0228 or 0301), different aerosol perturbations (0228$\_$NA2 or 0301$\_$NA2) have a different impact on the cloud and thermodynamical processes, in terms of magnitude or vertical location, as shown in the vertical profiles (gray and
orange curves in \Fig{vp_diff}) compared to those of 0228$\_$NA1 and 0301$\_$NA1, correspondingly.
The response evolves with time as well. The time evolution of
vertical-profiles difference for the February 28 case is shown in \Fig{vp_t_diff_0228}
. The aerosol effect on $\theta$ at the inversion
layer evolves from positive to negative and stays positive within
the boundary layer. The opposite behavior of aerosol effect on
$q_v$ and $q_c$ is observed. \Fig{vp_t_diff_0301} shows the one for the March 1
case. Unlike the February 28 case, the positive effect on $\theta$ and negative effect
on $q_v$ and $q_c$ of aerosols persist to the end of the
simulation and is more profound near the evolving
cloud top as discussed. $\Delta d\theta/dt$ and $\Delta d q_v/dt$
due to condensation shows the most distinct structure at the cloud top and the cloud base. 
The response of TKE vertical-profiles to aerosols
are more complicated. It exhibits strong spatial variations
for both February 28 and March 1 cases.
Overall, ACI has discernible but variable effects on the BL meteorology that depend on the BL structure and clouds.

\section{Evaluating cloud micro/macro-properties in ERA5 and MERRA-2}
\label{sec:comp}

In this section, we evaluate ERA5 and MERRA-2 reanalysis by comparing the LWP,
CFC, $N_c$, and $r_{\rm eff}$ from LES and GOES-16 to those from ERA5
and MERRA-2 reanalysis. Both ERA5 and MERRA-2 provide LWP as part of their
hourly single-level data. The CFC field of ERA5/MERRA-2 is defined as the
maximum cloud fraction in the vertical (up to 7 km) based on the model-level LWC
data with a threshold of ${\rm LWC}=0.02\, \rm{ g\, cm}^{-3}$ for cloud. Note
that ERA5 and MERRA-2 provide 1-hourly and 3-hourly data (instantaneous field)
for the CFC calculation, respectively.
The GOES-16 LWP is corrected using the
mean bias reported in \citet{painemal2012goes} and
\citet{painemal2021evaluation}.

\begin{figure*}[t!]\begin{center}
\includegraphics[width=\textwidth]{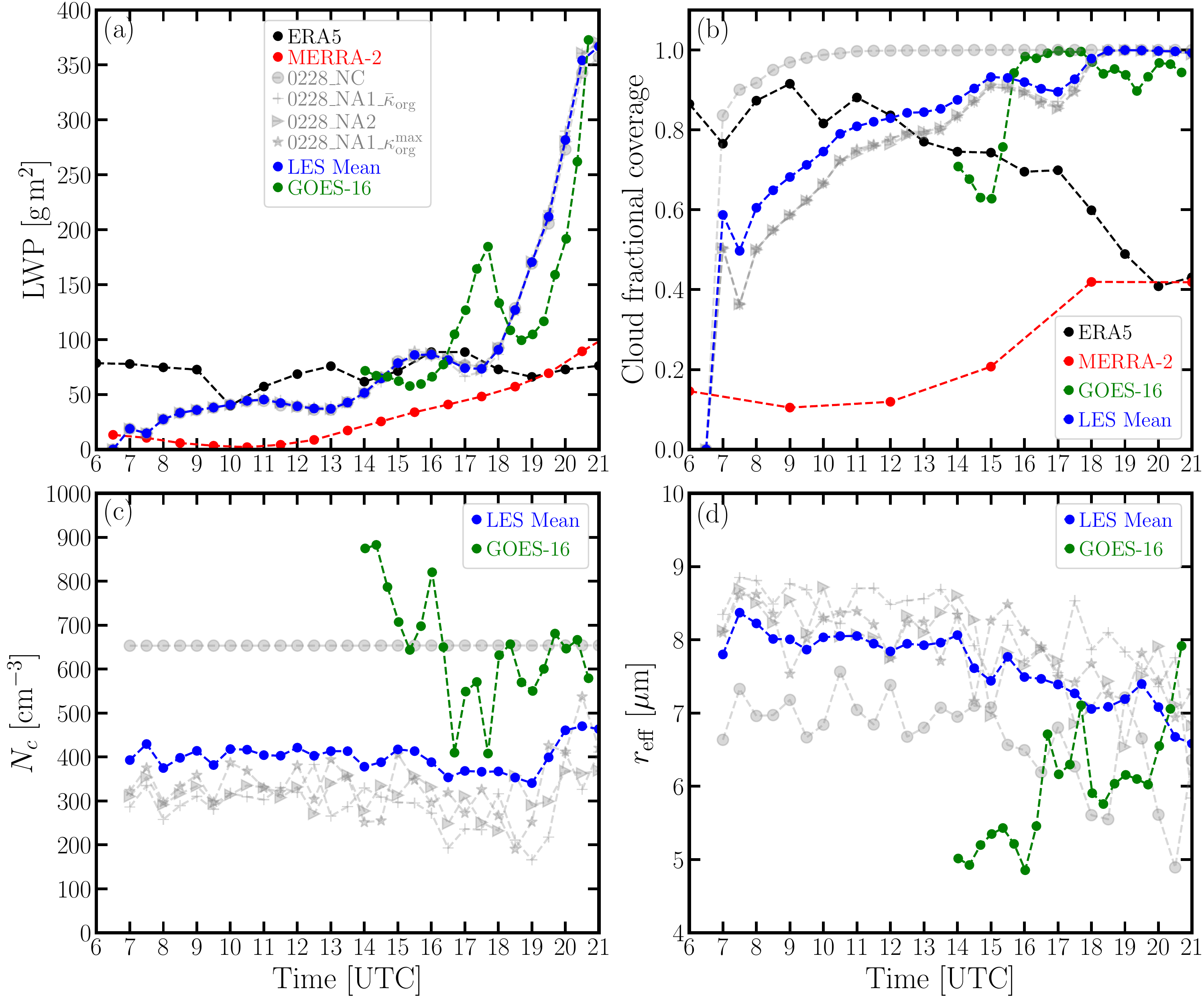}
\end{center}\caption{Comparison between the WRF-LES (blue, averaged over
all the four simulations), ERA5 (black), MERRA-2 (red), and GOES-16 (green) for the
February 28 case. The gray dashed-dotted lines represent individual simulations.
LWP is retrieved from hourly single-level (quantities
obtained from the model level) ERA5 and MERRA-2 (starting from 00:30 UTC). The
mean LWP is calculated by averaging over the dropsonde-measurement area. 
Both ERA5 (hourly) and MERRA-2 (3-hourly) provides Cloud Fractional Coverage (CFC) field at individual
model level. The time evolution of CFC is obtained by extracting the maximum
values of the CFC vertical-profiles below 7 km.  Vertical profiles are
obtained by sampling each layer conditionally with a threshold of ${\rm
LWC}=0.02\, \rm{ g\, cm}^{-3}$ for clouds.
%MERRA-2 CFC provides 3-hourly data with instantaneous output.
$N_c$ and $r_{\rm eff}$ are averaged over the cloud top (4 layers) from the WRF-LES ouput. 
ERA5, MERRA-2, and GOES-16 data are
averaged over the dropsonde area.
LWP, $N_c$, and $r_{\rm eff}$ from GOES-16 are filtered by cloud optical depth
$\ge 3$. GOES-16 CFC are obtained by normalizing the number of pixels with LWP
or IWP $\ge 0$ by the total number of pixels within the dropsonde measurement
area. To reduce the well-known systematic biases in GOES-16 retrievals, the GOES-16 LWP is
corrected by the mean of the lower $+10\, {\rm g\, m^{-2}}$ and upper $+19\, {\rm g\,
m^{-2}}$ bias bounds reported in \citet{painemal2012goes} and
\citet{painemal2021evaluation}. The GOES-16 $r_{\rm eff}$ ($N_c$ according to \Eq{eq:Nc}) are
corrected by the mean of lower $-2.4\, \mu{\rm m}$ and upper $-4.0\, \mu{\rm m}$
bias bounds according to \citet{painemal2012goes} and
\citet{painemal2021evaluation}. GOES-16 data before 14:00 UTC are excluded
because the retrievals are less reliable for low solar angles.
}
\label{lwp_cc_0228}
\end{figure*}

\begin{figure*}[t!]\begin{center}
\includegraphics[width=\textwidth]{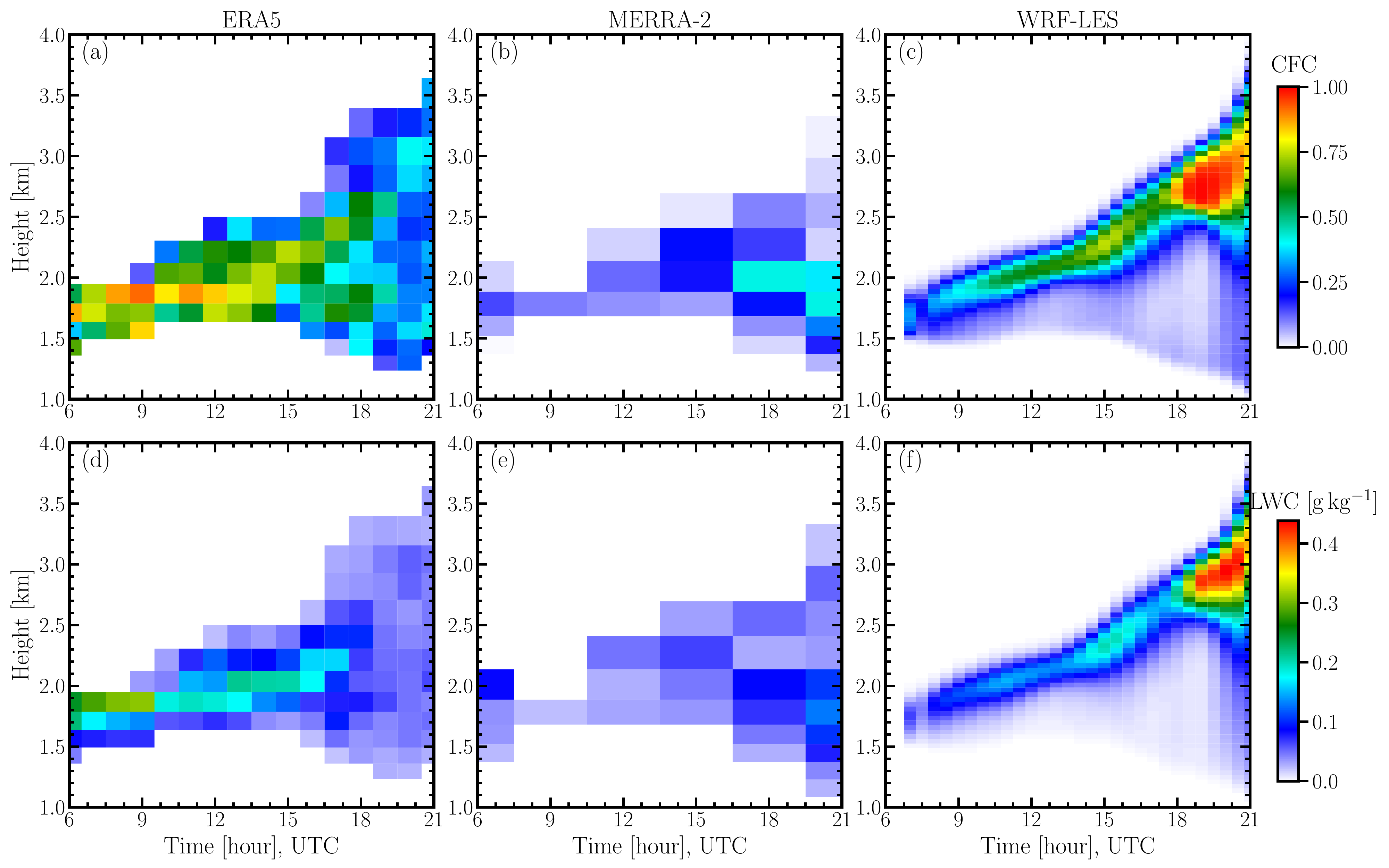}
\end{center}\caption{Evolution of vertical profile of cloud coverage and LWC up to 7 km for the
February 28 case. They are obtained by sampling each layer conditionally with a
threshold of ${\rm LWC}=0.02\, \rm{ g\, cm}^{-3}$.
ERA5 and MERRA-2 data are
averaged over the dropsonde area.
WRF-LES shows the averaged CFC from four simulations as in \Fig{lwp_cc_0228}.
}
\label{cc_ver0228}
\end{figure*}

\Fig{lwp_cc_0228} shows the comparison between LES, GOES-16, ERA5, and MERRA-2
for the February 28 case.
LWP (\Fig{lwp_cc_0228}(a)) and CFC (\Fig{lwp_cc_0228}(b)) retrieved from GOES-16
agrees reasonably well with the LES, which boosts our confidence to evaluate
the ERA5 and MERRA-2 using LES.
Compared to simulation $\febm$, ERA5 slightly
overestimates the LWP while MERRA-2 underestimates it. Nevertheless, both ERA5
and MERRA-2 capture the time evolution of LWP well. MERRA-2 has a significantly lower CFC
compared to the LES, while CFC has an opposite time evolution between
ERA5 and LES. We further compare
the evolution of vertical structure of LWC and
CFC between ERA5, MERRA-2, and LES. As shown in \Fig{cc_ver0228}, MERRA-2
underestimates the magnitude of CFC and LWC compared to the LES. Neither
MERRA-2 nor ERA5 captures the vertical structure of CFC and LWC of LES that has
an apparent peak near cloud top. 

We also compare $N_c$ and $r_{\rm eff}$ retrievals from GOES-16 with the LES
near the top of clouds. GOES-16 $N_c$ is derived from cloud effective radius in
$\mu$m and cloud optical depth $\tau$ under the adiabatic assumption. The
adiabatic lapse rate of condensation $\Gamma$ ($\rm{g\, m}^{-4}$) is estimated
from the cloud top temperature and pressure retrievals from GOES-16 (see section
\ref{app:Nc} for details of the estimation).
Since several studies have shown the presence of a systematic positive bias in satellite-based $r_{\rm eff}$ \citep{painemal2012goes, noble2015modis, zhang2017intercomparisons, painemal2021evaluation}, here we use values reported in \citet{painemal2012goes} and
\citet{painemal2021evaluation} for GOES-13 and GOES-16 to correct $r_{\rm eff}$ and, in turn, correct $N_c$ via \Eq{eq:Nc}. To be consistent with the sensitivity of satellite $r_{\rm eff}$ to the cloud uppermost layer, $N_c$ and $r_{\rm eff}$ from WRF-LES are averaged over the cloud top (4 layers, about 150 m). \Fig{lwp_cc_0228}(c) and \Fig{lwp_cc_0228}(d) show that $N_c$ and $r_{\rm eff}$ from WRF-LES agree reasonably well with those from GOES-16.

\begin{figure*}[t!]\begin{center}
\includegraphics[width=\textwidth]{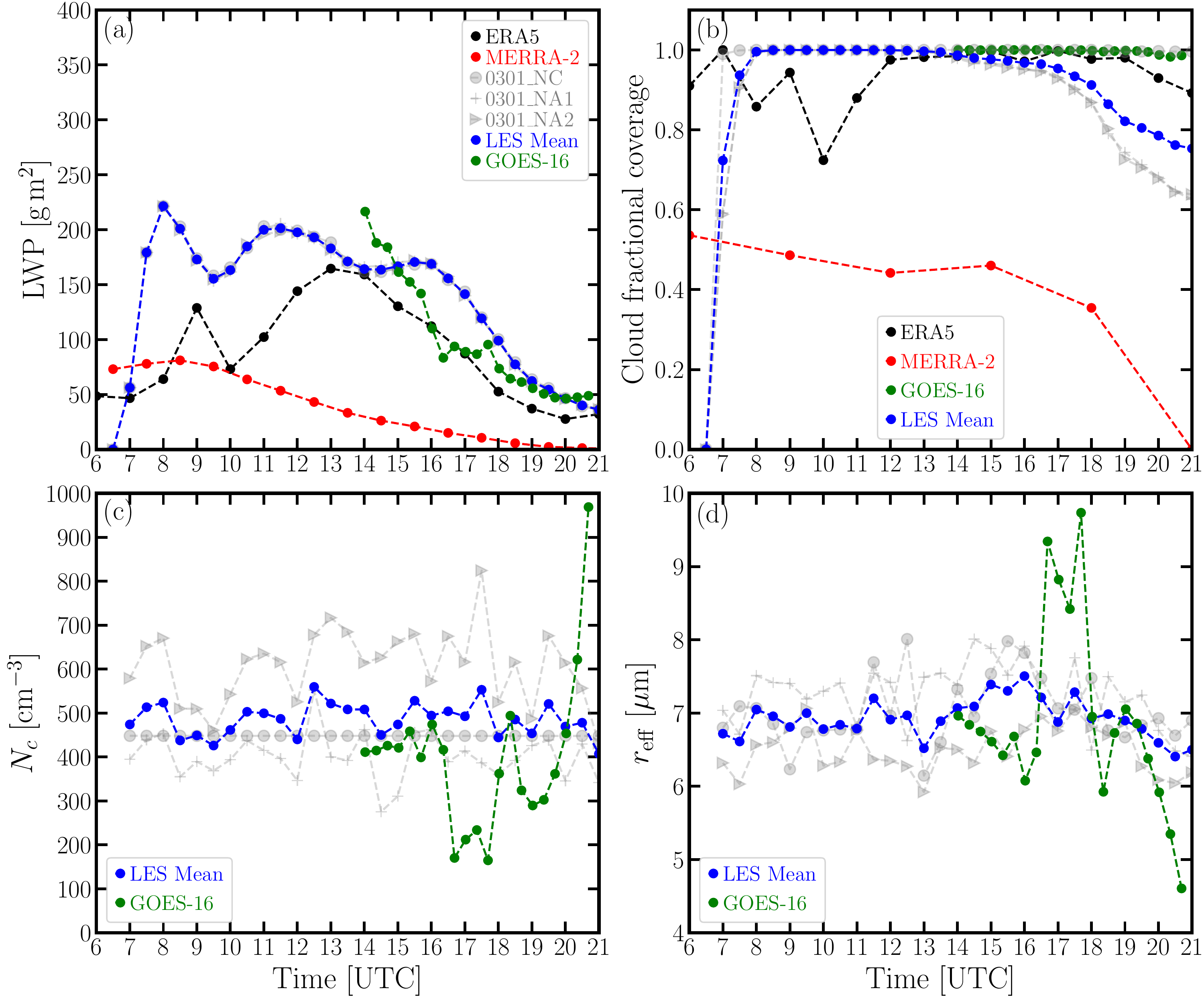}
\end{center}\caption{Same as \Fig{lwp_cc_0228} but for the March 1 case.
Blue dash-dotted lines represent the mean of three simulations.
}
\label{lwp_cc_0301}
\end{figure*}

\begin{figure*}[t!]\begin{center}
\includegraphics[width=\textwidth]{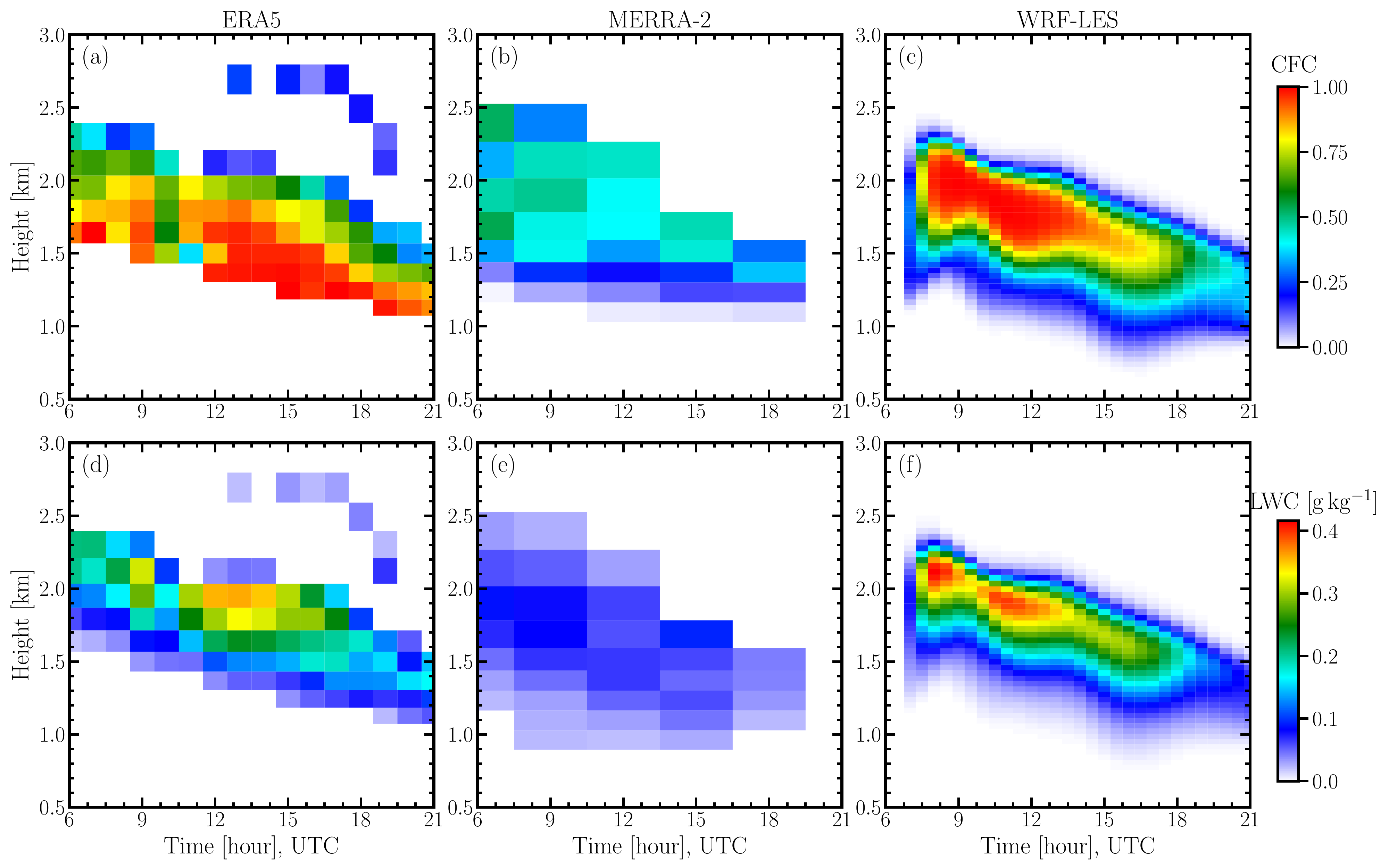}
\end{center}\caption{Same as \Fig{cc_ver0228} for the March 1 case.
WRF-LES shows the averaged CFC from three simulations as in \Fig{lwp_cc_0301}.
}
\label{cc_ver0301}
\end{figure*}

\Fig{lwp_cc_0301} shows the comparison between LES, GOES-16, and reanalysis for the March 1 case. The agreement in both magnitude and variation of LWP and CFC
between the LES and GOES-16 is reasonably good.
Compared to the LES and GOES-16, both MERRA-2 and ERA5 underestimate LWP. However ERA5
agrees better with the LES in the diurnal variation.
ERA5 agrees with LES in CFC while MERRA-2
underestimates CFC. The vertical profiles of LWC and CFC are shown in \Fig{cc_ver0301}.
Similar to the February 28 case, neither ERA5 nor MERRA-2
capture the vertical structure of CFC compared to LES.
Nevertheless, they exhibit comparable vertical structure of
LWC to the LES. ERA5 has larger CFC (nearly overcast condition)
near cloud base and a LWC maximum in the middle of cloud layers.
MERRA-2 has the lowest LWC and CFC at all heights. 
GOES-16 gives a similar $N_c$ and $r_{\rm eff}$ compared to LES at the cloud top. 

\section{Discussions and conclusion}
\label{sec:last}

Using the WRF-LES, we have simulated two cold-air outbreak (CAO) cases (28 February and 1
March, 2020) observed during the ACTIVATE campaign in the Western North Atlantic
Ocean. The aerosol-meteorology-cloud interaction (AMCI) during the
two CAO events are examined by comparing WRF-LES, measurements, satellite retrievals,
and reanalysis. 

Aerosol size distributions measured from airborne aerosol-sampling
are prescribed to the LES experiments, which are compared to previous experiments with prescribed cloud droplet number concentrations \citep{2021arXiv210706193L}. Cloud properties from the LES experiments are then validated
against the FCDP in-situ measurements. Log-normal aerosol size distributions are fitted
from data acquired during two below-cloud-base (BCB) flight legs for each case. 
Bulk aerosol hygroscopicity estimated from mass of individual aerosol components measured by the Aerodyne Mass Spectrometer
is used in the LES. For organic species, we need to make assumptions in the component hygroscopicity.
To examine the effect of aerosols, we compared LES results with prescribed
aerosol size distributions from different flight legs and/or with different assumptions in the estimation of bulk hygroscopicity (denoted as NA-LES) to the ones with constant cloud droplet
number concentration $N_c$ (denoted as NC-LES) obtained from the FCDP sampling. For the
February 28 case, vertical profiles of LWC from the two NA-LES are in good agreement. However, $N_c$ ($r_{\rm eff}$) from the NC-LES
is larger (smaller) than that from NA-LES. This is speculated to be because the aerosol sampling
during the two BCB legs were not co-located very well with the FCDP sampling in
space. For the March 1 case, LWC, $N_c$, and $r_{\rm eff}$ from NC-LES agree
better with those from NA-LES using aerosol size distributions derived from the
BCB flight leg that co-located well in space with the FCDP cloud droplet sampling.      
Our LES-measurement comparison also demonstrates a strong spatial and temporal
variation of aerosol particle distributions during the CAO events, which adds challenges to the LES modeling and validation.
By comparing LES to the measurements during flights on the same day
but at a later time, we show that the LES captures the diurnal variation
of cloud properties for both cases.
Overall, the good agreement between WRF-LES and measurements gives us confidence to use the LES results to study the AMCI and evaluate reanalysis for the two CAO events.

Aerosol effects on cloud micro-physical and macro-physical 
properties (i.e., $N_c$, and $r_{\rm eff}$, LWP and CFC) are investigated by comparing
LES with aerosol size distributions from different BCB flight legs as input.
Our use of measured aerosol size distributions to study the aerosol effects
on cloud properties better represents aerosol perturbations in the study domain, which is more
realistic compared to other LES studies with prognostic $N_c$ but idealized aerosol size distribution.
More importantly, LES with the same configuration and boundary/surface forcings but different aerosol perturbations
allows us to disentangle the aerosol effect on clouds from the meteorological effect,
which is challenging in understanding ACI \citep{stevens2009untangling}.
For the February 28 case, $N_c$ and $r_{\rm eff}$ are influenced by aerosols
via aerosol hygroscopicity $\bar{\kappa}$ and aerosol size distributions.
With the same aerosol size distributions, increasing $\bar{\kappa}$ from 0.313 to 0.392 leads to $5.2\%$ increase ($3.4\%$ decrease) of $N_c$ ($r_{\rm eff}$).
The top-of-the-atmosphere short wave radiation
$\rm{SW_{TOA}}$ decreases by $1.13\, \rm{W\, m}^{-2}$.
LWP (IWP) increases by $3.2\%$ ($13.1\%$).
The CFC only decreases by $0.4\%$.
Even though aerosol size distributions from two BCB
legs are quite similar, a difference of $1.11\, \rm{W\, m}^{-2}$
in $\rm{SW_{TOA}}$ is observed for the two simulations.
For the March 1 case, $\rm{SW_{TOA}}$ only changes by
$0.72\, \rm{W\, m}^{-2}$ when
$N_a$ differs by a factor of about 2.6
between two BCB flight legs.
The effect of $N_a$ on LWP and CFC is negligible.

As we aim to eventually evaluate and improve ACI processes in Earth
system models using LES experiments informed by the ACTIVATE observations, we have also
compared our LES results to satellite retrievals (GOES-16) and reanalysis products, such as ERA5 and MERRA-2.
For the February 28 case, LES and GOES-16 agree reasonably well in LWP and CFC.
ERA5 captures the LWP compared with LES while MERRA-2 slightly underestimates LWP.
Time evolution and the magnitude of CFC among LES, ERA5, and
MERRA-2 are quite different.
Both ERA5 and MERRA-2 fail to capture the vertical structure of LWC and CFC compared to the LES.
For the March 1 case, LES agree well with GOES-16 in LWP and CFC.
Both ERA5 and MERRA-2 underestimate the LWP compared to
LES, even though the time evolution of LWP exhibit a similar trend between ERA5 and
the LES. ERA5 and LES agree well in the magnitude of CFC while MERRA-2 largely
underestimates it.
Similar to the February 28 case, the vertical structure
of LWP and CFC from MERRA-2 and ERA5 are quite different from the LES.
We have also validated $N_c$ and $r_{\rm eff}$ from the LES against the GOES-16
retrievals. For both cases, $N_c$ and $r_{\rm eff}$ from the LES agrees well with those from GOES-16. 

To use the LES results to evaluate AMCI in Earth system models, 
the next step will be to simulate these two cases using single-column configuration of Earth system models driven by the same boundary and surface forcings obtained from ACTIVATE measurements and reanalysis products.

\acknowledgments

This work was supported through the
ACTIVATE Earth Venture Suborbital-3 (EVS-3) investigation, which is
funded by NASA’s Earth Science Division and managed through the
Earth System Science Pathfinder Program Office.
C.V. and S.K. thank funding by the DFG within the SPP 1294 HALO and the TRR301-1 TP change.
The Pacific Northwest National Laboratory (PNNL) is
operated for the U.S. Department of Energy by Battelle Memorial Institute
under contract DE-AC05-76RLO1830.
The source code used for the simulations of this study, the
Weather Research and Forecasting (WRF) model,
is freely available on \url{https://github.com/wrf-model/WRF}.
The simulations were performed using resources available through
Research Computing at PNNL.

\datastatement

ACTIVATE Data are publicly available at: \url{http://doi.org/10.5067/SUBORBITAL/ACTIVATE/DATA001}

\appendix

\section{Validation of log-normal fitting of aerosol size distributions}

\begin{table*}[t!]
\caption{Fitted parameters of the aerosol size distribution
for the February 28 case shown in \Fig{dNdlnD_60221_60423_3modes_0228}.
The percentage error (PE) is defined as 
${\rm PE} = (\bar{N}_{\rm fit}-\bar{N}_{\rm a})/\bar{N}_{\rm a}\times 100\%$.
} 
\centering
\setlength{\tabcolsep}{1pt}
\begin{tabular}{|c|c|c|c|c|c|c|c|c|c|c|c|c|c|}
\hline
  \multirow{2}{*}{BCB leg} &\multirow{2}{*}{Time, UTC} & \multicolumn{3}{c|}{$N\, ({\rm cm}^{-3})$} & \multicolumn{3}{c|}{$\mu$ (nm)} & \multicolumn{3}{c|}{$\sigma$} & \multirow{2}{*}{$\bar{N}_{\rm a}\, ({\rm cm}^{-3})$} & \multirow{2}{*}{$\bar{N}_{\rm fit}\, ({\rm cm}^{-3})$} & \multirow{2}{*}{PE} \\
\cline{3-11}
&  & $N_1$ & $N_2$ & $N_3$ & $\mu_1$& $\mu_2$& $\mu_3$& $\sigma_1$& $\sigma_2$ & $\sigma_3$ & & & \\ 
\cline{3-11}
\hline
BCB1 & 15:46:53-15:54:54 & 4222 & 994 & 198 & 26.7 & 64.9 & 144.5 &1.46 &1.38 & 1.51 & 5593 & 5384 & $-3.7\%$ \\ 
\hline
BCB2 & 16:43:41-16:47:03 & 3757 & 723 & 219 & 33.0  & 63.0 & 173.7 & 1.49 & 1.49 & 1.40 & 5364 & 4690 & $-12.6\%$ \\ 
\hline
\multicolumn{14}{p{0.3\textwidth}}{}
\end{tabular}
\label{tab:fit}
\end{table*}

\begin{table*}[t!]
\caption{Fitted parameters of the aerosol size distribution
for the March 1 case shown in \Fig{dNdlnD_53602_54105_2modes_0301}.
} 
\centering
\setlength{\tabcolsep}{1pt}
\begin{tabular}{|c|c|c|c|c|c|c|c|c|c|c|c|c|c|}
\hline
  \multirow{2}{*}{BCB leg} &\multirow{2}{*}{Time, UTC}& \multicolumn{3}{c|}{$N\, ({\rm cm}^{-3})$} & \multicolumn{3}{c|}{$\mu$ (nm)} & \multicolumn{3}{c|}{$\sigma$} & \multirow{2}{*}{$\bar{N}_{\rm a}\, ({\rm cm}^{-3})$} & \multirow{2}{*}{$\bar{N}_{\rm fit}\, ({\rm cm}^{-3})$} & \multirow{2}{*}{PE} \\
\cline{3-11}
&  & $N_1$ & $N_2$ & $N_3$ & $\mu_1$& $\mu_2$& $\mu_3$& $\sigma_1$& $\sigma_2$ & $\sigma_3$ & & & \\ 
\cline{3-11}
\hline
BCB1 &14:53:22-15:01:45 & 940 & 645 & --  & 22.4 & 104.2 & --  &1.51 &1.47 &  -- & 1434 & 1479 & $3.1\%$ \\ 
\hline
BCB2 &15:51:21-15:55:06 & 996 & 1192 & 1118 & 19.0  & 30.2 & 102.3 & 1.49 & 1.31 & 1.51 & 3100 & 3139 & $1.3\%$ \\ 
\hline
\multicolumn{14}{p{0.3\textwidth}}{}
\end{tabular}
\label{tab:fit0301}
\end{table*}

\begin{figure*}[t!]\begin{center}
\begin{overpic}[width=0.48\textwidth]{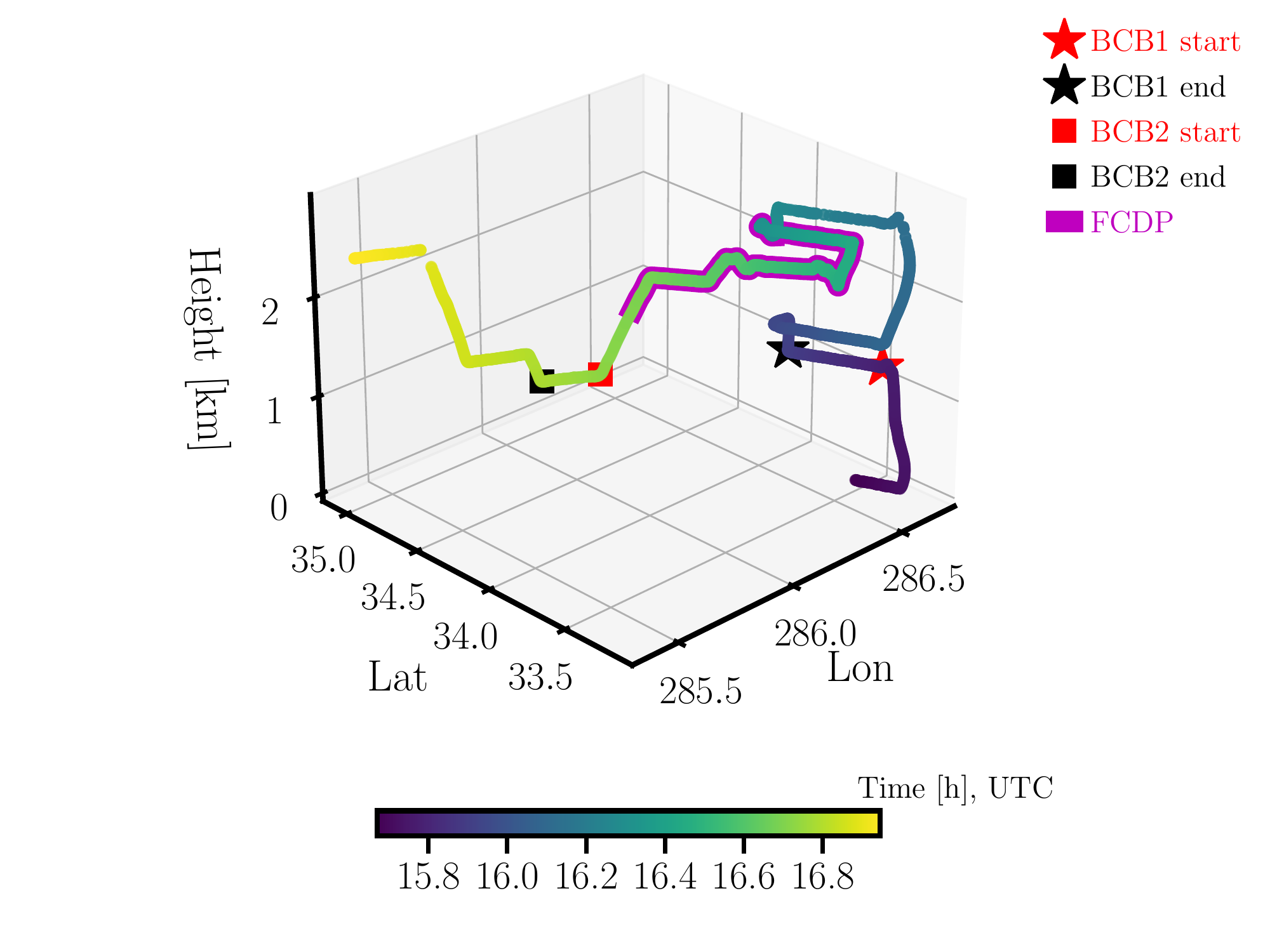}\put(15,70){(a)}\end{overpic}
\begin{overpic}[width=0.48\textwidth]{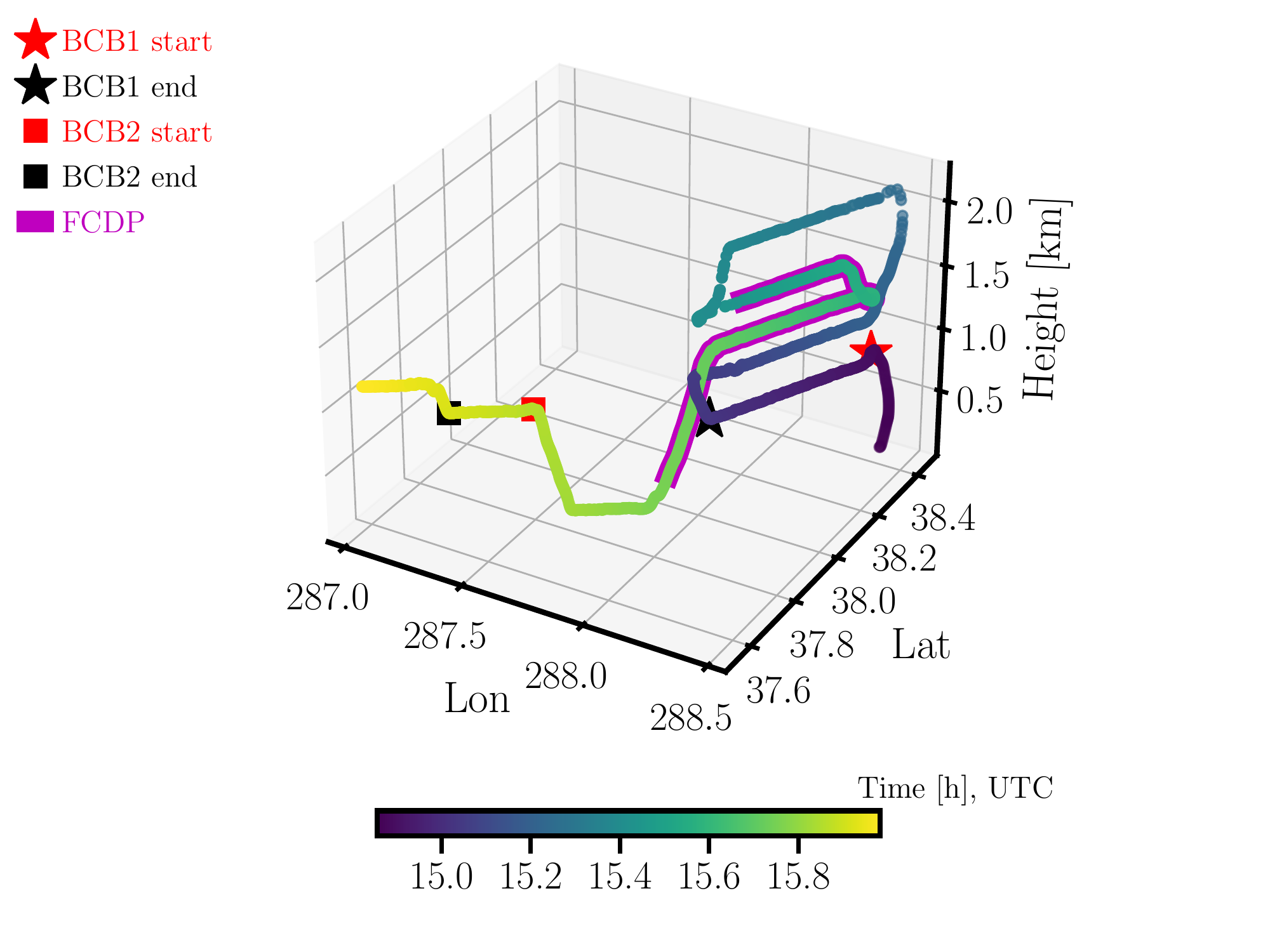}\put(70,70){(b)}\end{overpic}
\end{center}\caption{Flight trajectory during the FCDP
measurement for (a) the February 28
and (b) March 1 cases. Red and black stars (squares) represent the start and end of the BCB1 (BCB2) flight leg, respectively.
Trajectories with most
dense FCDP-sampling (16.3-16.7 and 15.45-15.75 UTC for the February 28 and March 1 cases, respectively) are highlighted as thick magenta curves.   
}
\label{FCDP_traj}
\end{figure*}

\begin{figure*}[t!]\begin{center}
\begin{overpic}[width=\textwidth]{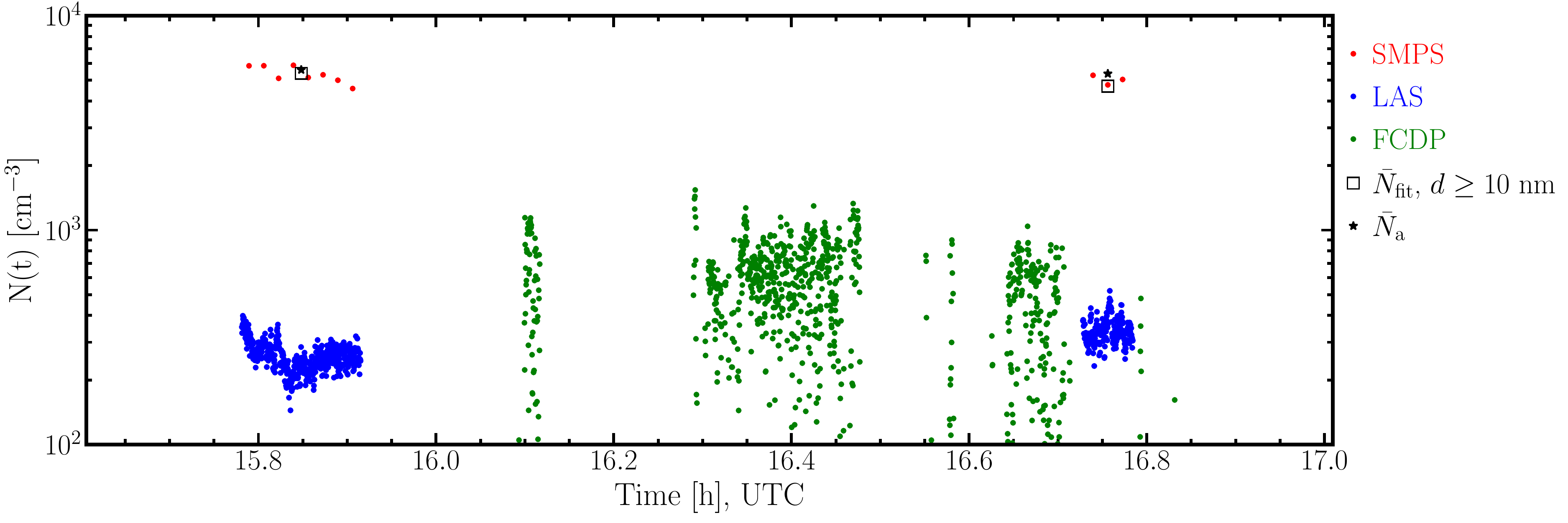}\put(6.5,30){(a)}\end{overpic}
\begin{overpic}[width=\textwidth]{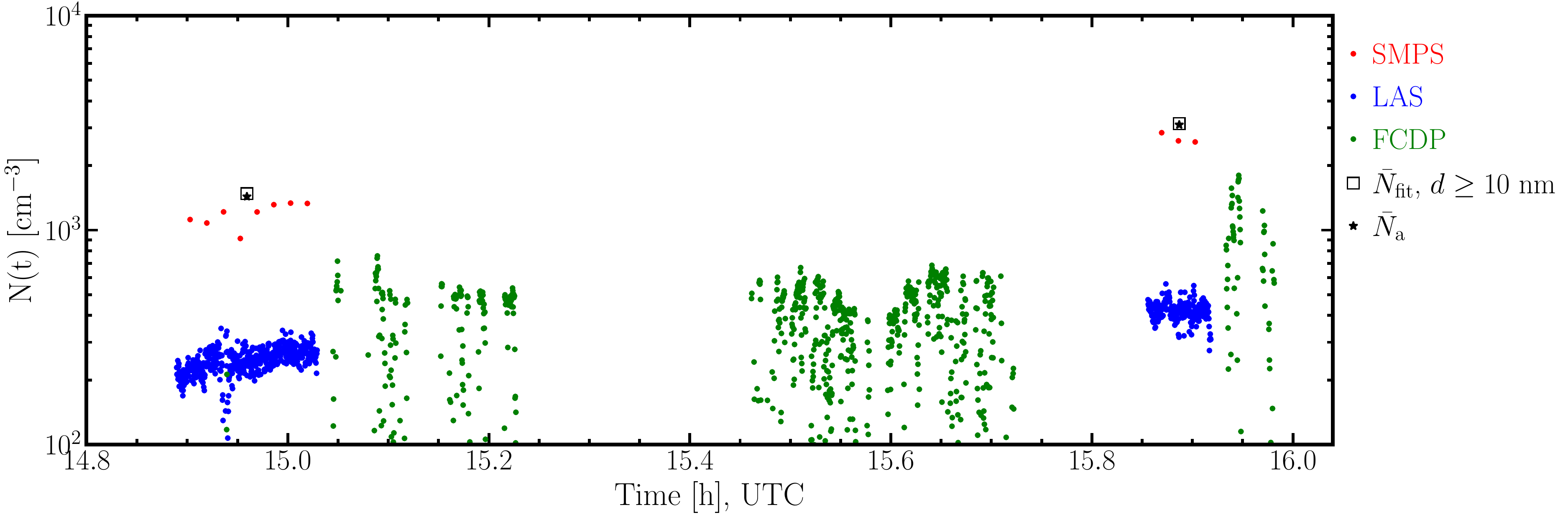}\put(6.5,30){(b)}\end{overpic}
\end{center}\caption{Comparison between
%$N_{\rm CPC}$ (particles with $d\ge10\, {\rm nm}$, black dots),
$\bar{N}_{\rm a}$ (black star), $\bar{N}_{\rm fit}$ (black square),
and $N_{\rm FCDP}$ (green dots) for (a) the February 28 and (b) March 1 case. Red and blue dots represent $N_a$ from SMPS and LAS
measurements, respectively.
}
\label{cpc_0228}
\end{figure*}

The 3-D flight trajectories of the BCB legs for both cases are shown in \Fig{FCDP_traj}. 
\Fig{cpc_0228} shows time series of measured particle
number concentration of aerosols (SMPS and LAS)
%, condensation particles (CPC), 
and cloud droplets (FCDP) during the two BCB legs
for both cases.
The SMPS measurements are free from spikes, indicative of
no cloud-artifacts (e.g., the cloud shattering), during
the measurements. 
The validation of the log-normal fitting of aerosol size distributions
is discussed in section~\ref{sec:obs}\ref{sec:aerosol}.

\begin{table*}[t!]
\caption{Time-averaged mass concentration $\overline{m_i}$ from the AMS measurement sampled during
BCB flight legs for the February 28 (0228) and March 1 (0301) cases.
NaCl is not efficiently sampled by AMS because it is refractory (i.e., not volatile at 600 Pa), and therefore the Cl mass is likely not representative of NaCl mass.
It is justified by the lack of coarse-mode number concentration in \Fig{dNdlnD_60221_60423_3modes_0228} and
\Fig{dNdlnD_53602_54105_2modes_0301}.
} 
\centering
\setlength{\tabcolsep}{1pt}
\begin{tabular}{|c|c|c|c|c|c|}
\hline
Component & & organic & Sulfate (${\rm SO}_4^{2-}$) &
Nitrate (${\rm NO}_3^{-}$) & Ammonium (${\rm NH}_4^{+}$) \\
\hline
\multirow{2}{*}{0228} & BCB1 & $54.5\%$ & $23.4\%$ & $10.8\%$ & $10.0\%$\\
 & BCB2 & $48.6\%$ & $25.9\%$ & $10.8\%$ & $13.9\%$\\
\hline 
\multirow{2}{*}{0301}  &BCB1 & $27.6\%$ & $46.6\%$ & $3.4\%$ & $21.3\%$ \\
  &BCB2 & $24.9\%$ & $28.5\%$ & $23.2\%$ & $22.6\%$ \\
\hline
%\multicolumn{6}{p{0.3\textwidth}}{}
\end{tabular}
\label{tab:m}
\end{table*}

\begin{table*}[t!]
\caption{$\bar{\kappa}$ (time-averaged $\kappa$) calculated according
to \Eq{eq:kappa} with AMS-measured $\overline{m_i}$ as input listed
in \Tab{tab:m}.
$\kappa_i$ is adopted from Table 1 of \citet{petters2007single} for
both the non-organic components and the organic one.
The mass of ${\rm NH}_4^{+}$ is divided to ${\rm (NH_4)_2 SO_4}$
and ${\rm NH_4NO_3}$ by its molecular proportion
assuming both sulfate and nitrate are fully neutralized
as ${\rm (NH_4)_2 SO_4}$ and ${\rm NH_4NO_3}$.
Taking the upper limit of the kappa value for the organic aerosols
as $\kappa_{\rm org}=0.229$ during the BCB1 sampling
for the February 28 case, we get $\bar{\kappa}=0.392$.
} 
\centering
\setlength{\tabcolsep}{1pt}
\begin{tabular}{|c|c|c|c|c|c|}
\hline
  Component & & organic & ${\rm (NH_4)_2 SO_4}$ &
${\rm NH_4NO_3}$  & $\bar{\kappa}$ \\
\hline
  $\rho_{i}\, ({\rm g\, cm}^{-3})$ & & 1.35 & 1.77 & 1.72 & \\
\hline
  $\kappa_i$ & & 0.1 & 0.61 & 0.67 & \\
\hline
  \multirow{2}{*}{0228} & BCB1 & $54.5\%$ & $30.0\%$ & $14.1\%$ & 0.313  \\
  & BCB2 & $48.6\%$ & $35.1\%$ & $15.5\%$ & 0.341 \\
  \hline
  \multirow{2}{*}{0301} & BCB1 & $27.6\%$ & $60.8\%$ & $10.5\%$ & 0.451\\
  &BCB2 & $24.9\%$ & $43.5\%$ & $30.7\%$ & 0.479\\
\hline
\end{tabular}
\label{tab:kappa}
\end{table*}

\section{Diurnal cycle of cloud properties}

\begin{figure*}[t!]\begin{center}
\begin{overpic}[width=\textwidth]{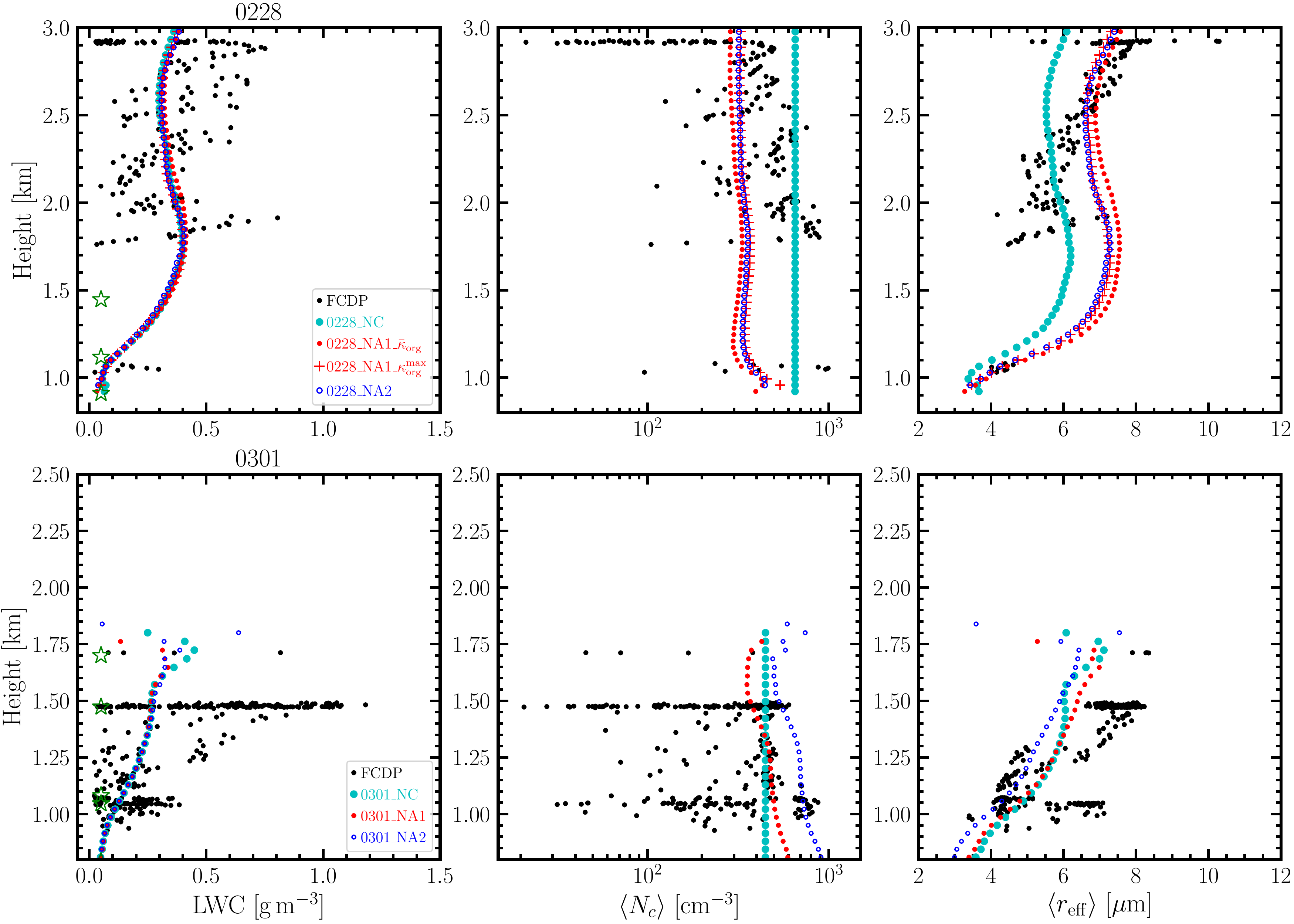}\put(-1,68.5){(a)}\put(-1,34){(b)}\end{overpic}
\end{center}\caption{Same as \Fig{FCDP_WRF_comp} but for
the flights (a) from 20:47:07 to 21:10:58 UTC and
(b) from 19:43:30 to 20:13:44 UTC
for the February 28 and March 1 cases, respectively.
}
\label{FCDP_WRF_comp_L2}
\end{figure*}

\begin{figure*}[t!]\begin{center}
\includegraphics[width=\textwidth]{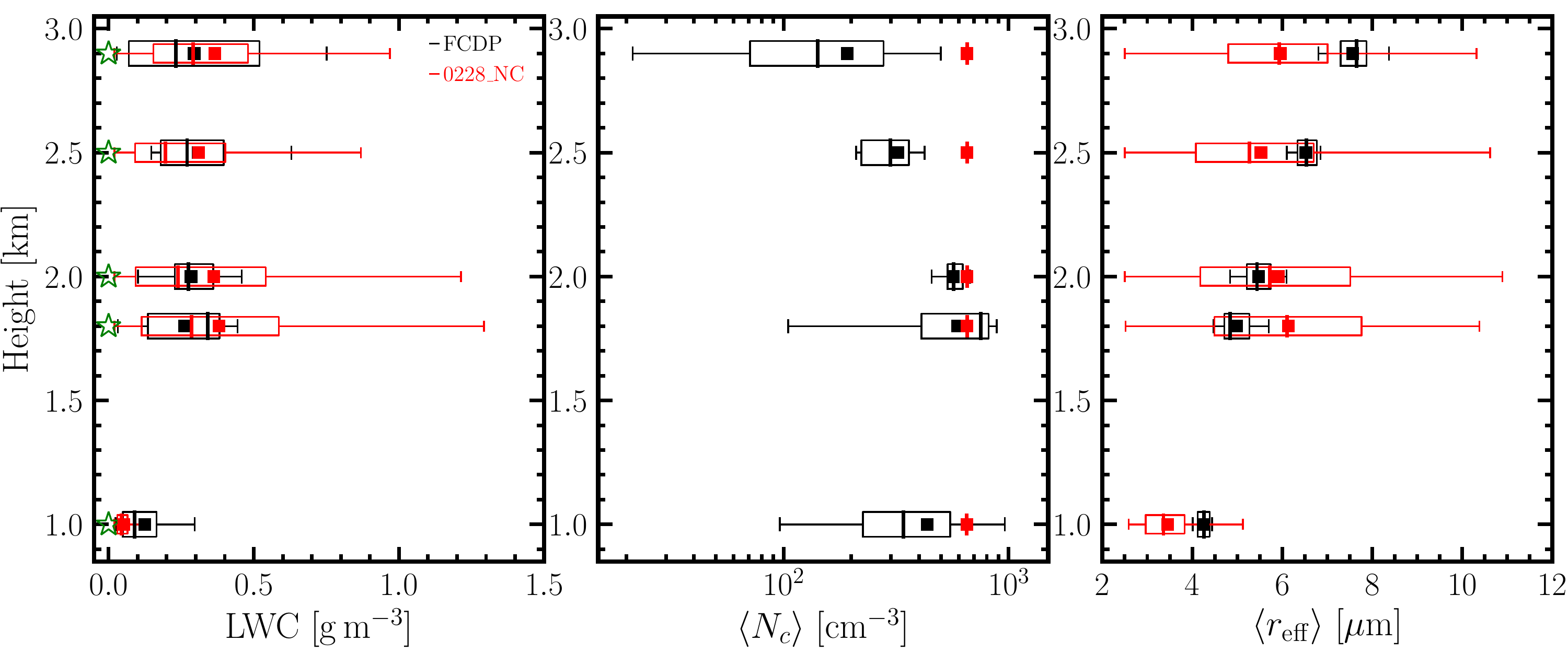}
\includegraphics[width=\textwidth]{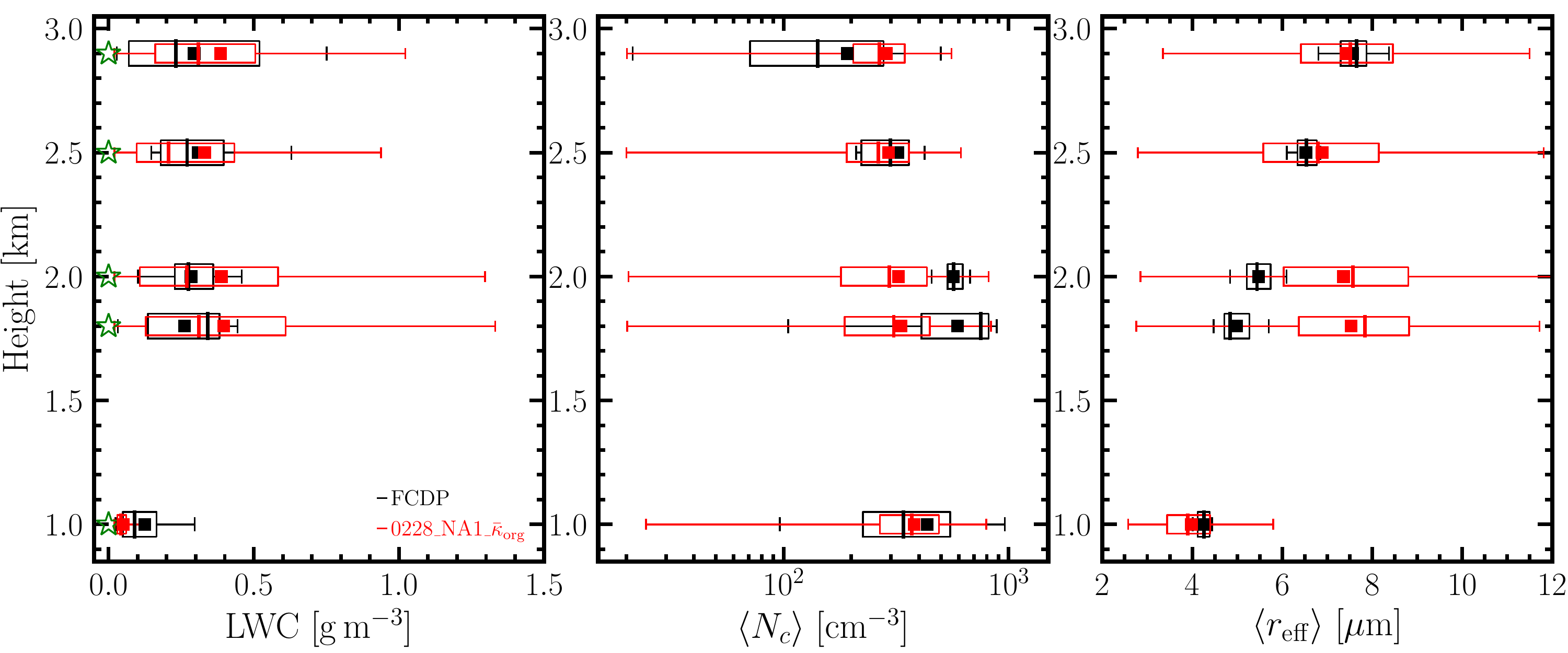}
\end{center}\caption{Same as \Fig{FCDP_0228} but for
the measurements from 20:47:07 to 21:10:58 UTC.
}
\label{FCDP_L2_0228}
\end{figure*}

\begin{figure*}[t!]\begin{center}
\includegraphics[width=\textwidth]{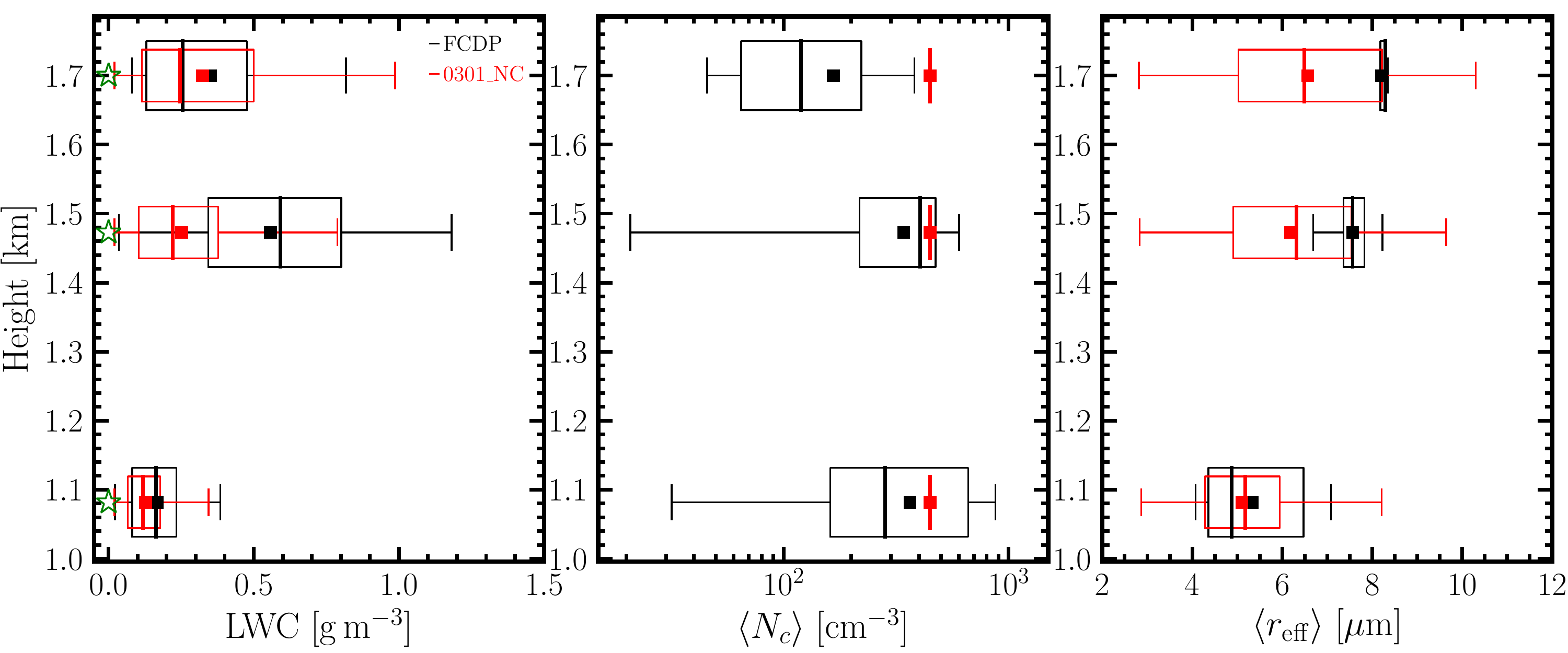}
\includegraphics[width=\textwidth]{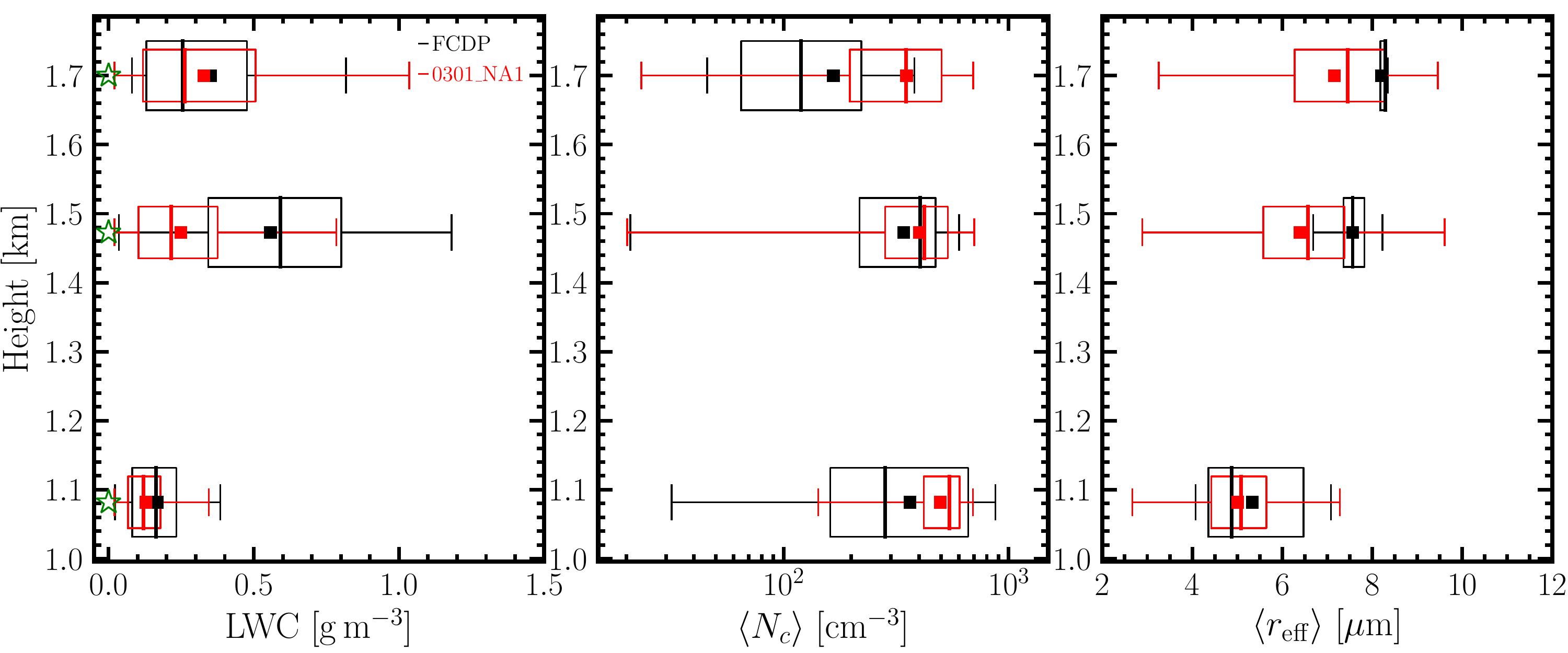}
\end{center}\caption{Same as \Fig{FCDP_0228_NA} but for
the measurements from 19:43:30 to 20:13:44 UTC.
}
\label{FCDP_L2_0301}
\end{figure*}

\Fig{FCDP_WRF_comp_L2} shows comparison between
the FCDP measurement and LES for the two cases
during the afternoon flights. The corresponding statistics are
shown in \Fig{FCDP_L2_0228} and \Fig{FCDP_L2_0301}
respectively.

\section{Ice}

\begin{figure*}\begin{center}
\includegraphics[width=\textwidth]{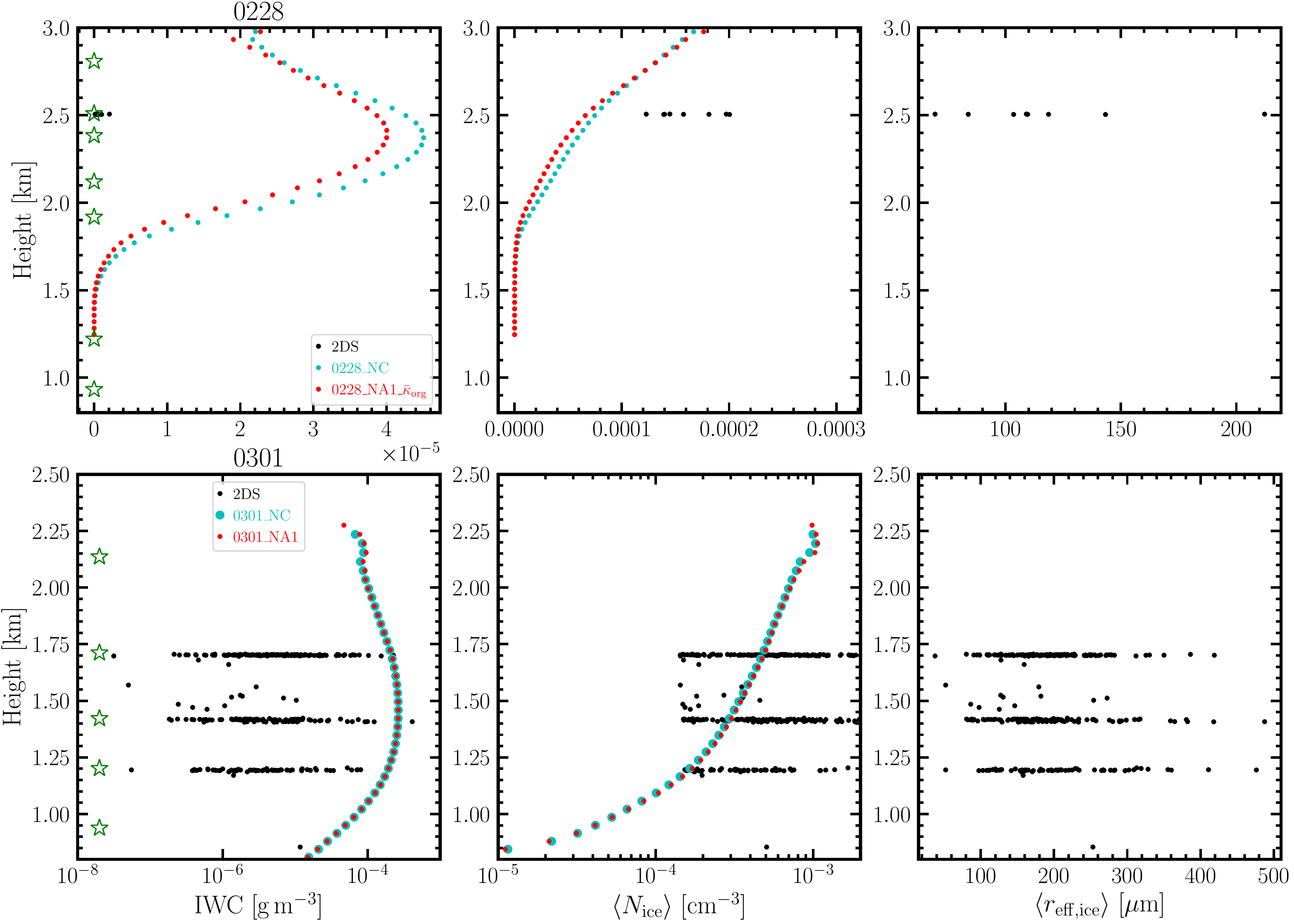}
\end{center}\caption{Comparison between the WRF-LES and the 2DS measurement of ice particles.
%The same criterion and procedure as in \Fig{FCDP_WRF_comp} are used.
The IWC from WRF-LES only includes ice particles.
WRF-LES does not output $r_{\rm eff, ice}$.
The $r_{\rm eff, ice}$ is the largest radius from the 2DS measurement.
}
\label{2DS_WRF_ice}
\end{figure*}

\begin{figure}\begin{center}
\includegraphics[width=0.5\textwidth]{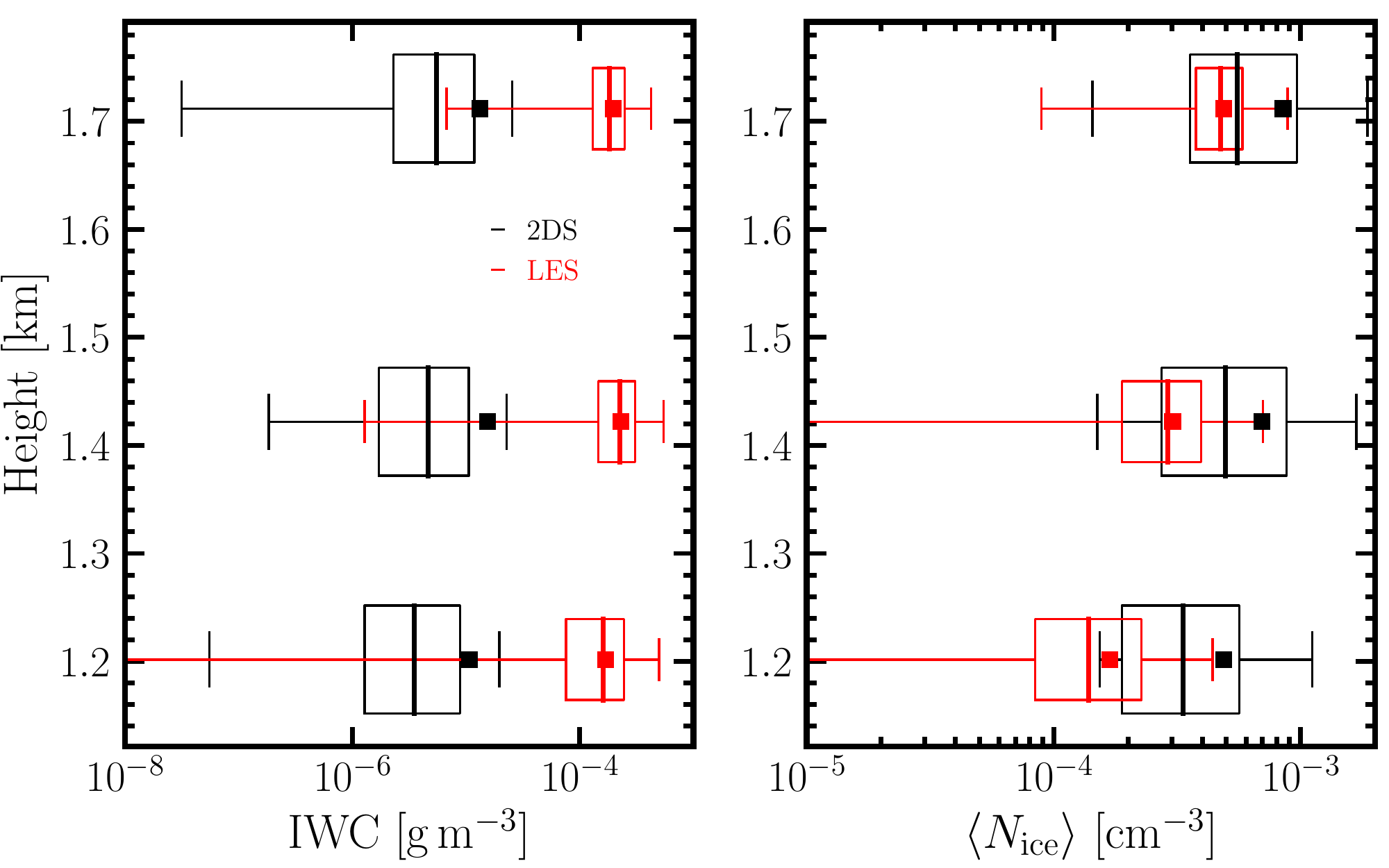}
\end{center}\caption{Corresponding statistics of \Fig{2DS_WRF_ice}.
2DS measurement for the March 1 case for prescribed $N_a$.
The same criterion and procedure for the box-and-whisker
plot as in \Fig{FCDP_0228} are used.
The green stars mark the altitude of flight legs.
}
\label{2DS_0301_NA}
\end{figure}

\Fig{2DS_WRF_ice} shows the vertical profiles of IWC,
$\langle N_{\rm ice}\rangle$, and $\langle r_{\rm eff, ice} \rangle$.
Hardly any ice was observed for the February 28 case.
For the March 1 case, $\langle N_{\rm ice}\rangle$ from WRF-LES
agrees well with the 2DS measurement as shown in \Fig{2DS_0301_NA}.

\section{Vertical profiles}

\begin{figure*}[t!]\begin{center}
\includegraphics[width=\textwidth]{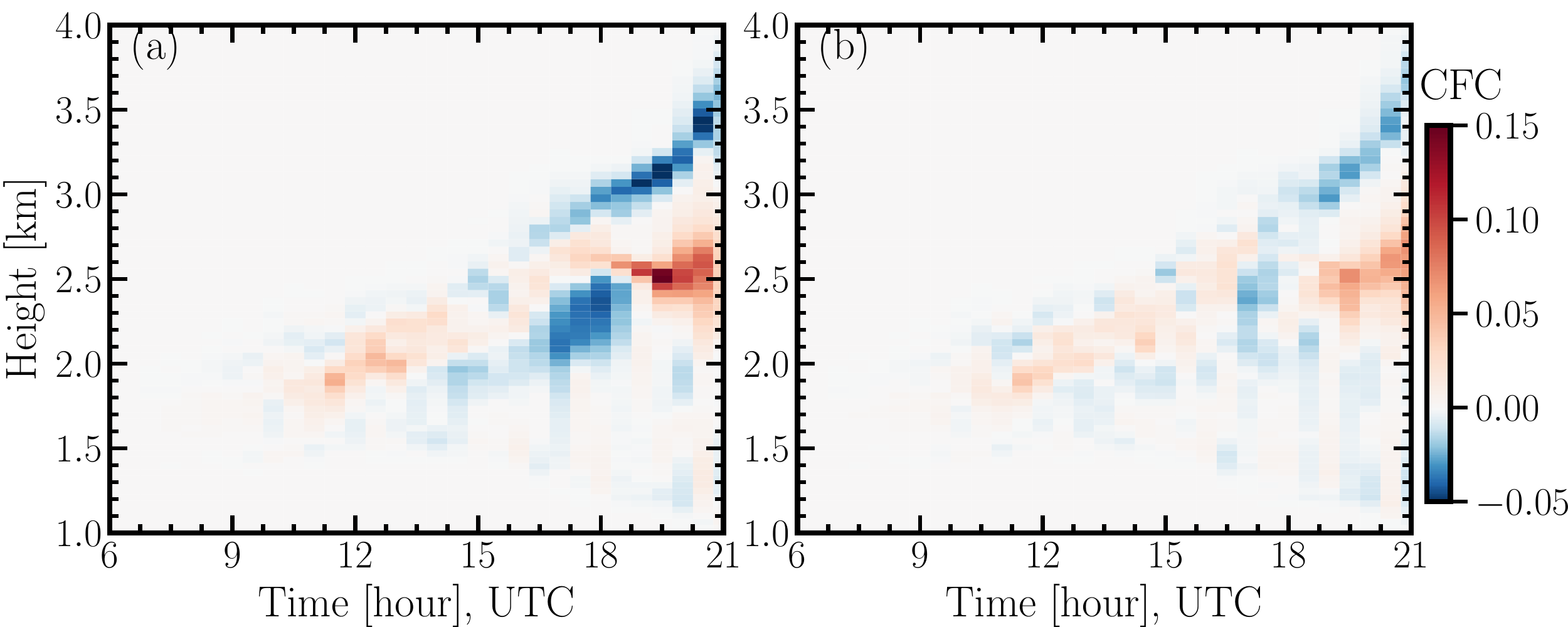}
\end{center}\caption{Evolution of difference of CFC profiles between simulation
(a) $\febm$ and 0228$\_$NC and between (b) 0228$\_$NA2 and 0228$\_$NC averaged during the
dropsonde measurement time 16:00-17:00 UTC for the February 28 case. 
}
\label{cfc0228}
\end{figure*}

\begin{figure*}[t!]\begin{center}
\includegraphics[width=\textwidth]{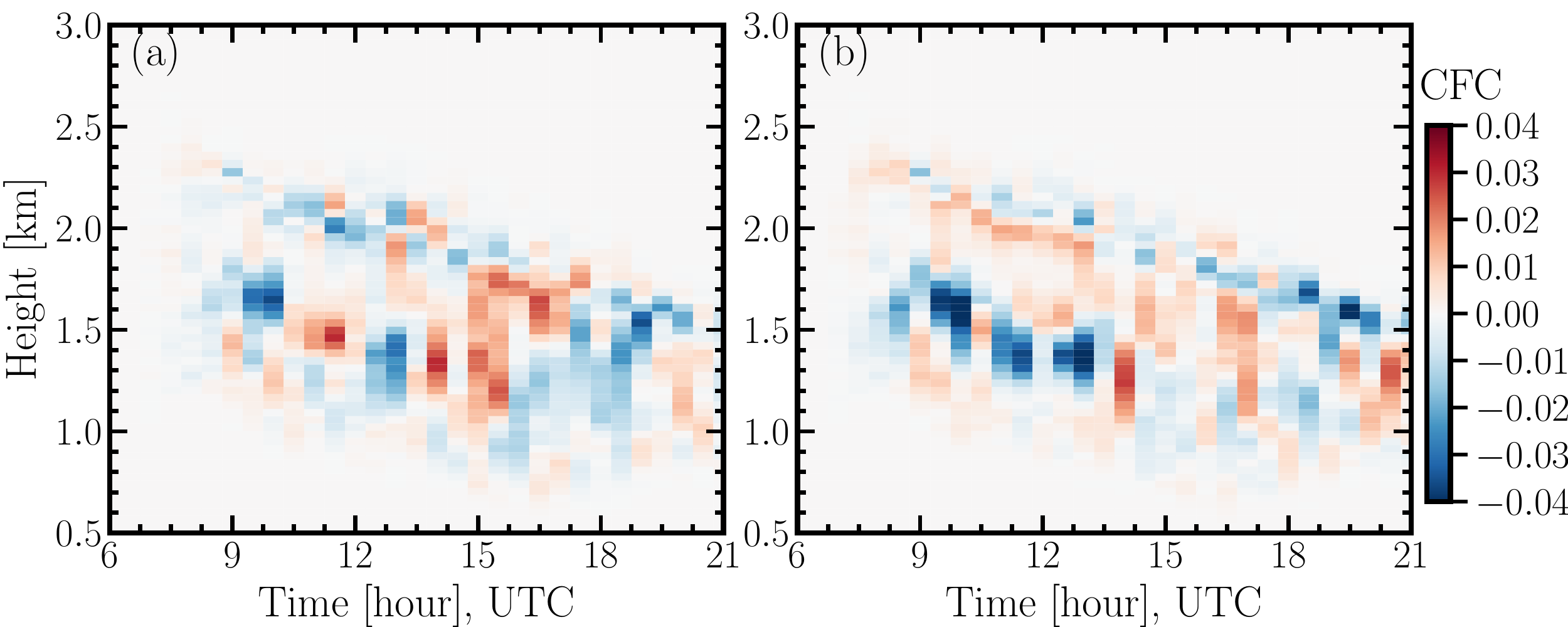}
\end{center}\caption{Same as \Fig{cfc0228} but for the March 1 case:
difference between (a) 0301$\_$NA1 and 0301$\_$NC and between (b) 0301$\_$NA2 and
0301$\_$NC averaged over 15:00-16:00 UTC. 
}
\label{cfc0301}
\end{figure*}

\Fig{cfc0228} and \Fig{cfc0301} shows the evolution of difference of
CFC profiles between NA and NC simulations for the February 28 and March 1
cases, respectively.

\begin{figure*}[t!]\begin{center}
\includegraphics[width=\textwidth]{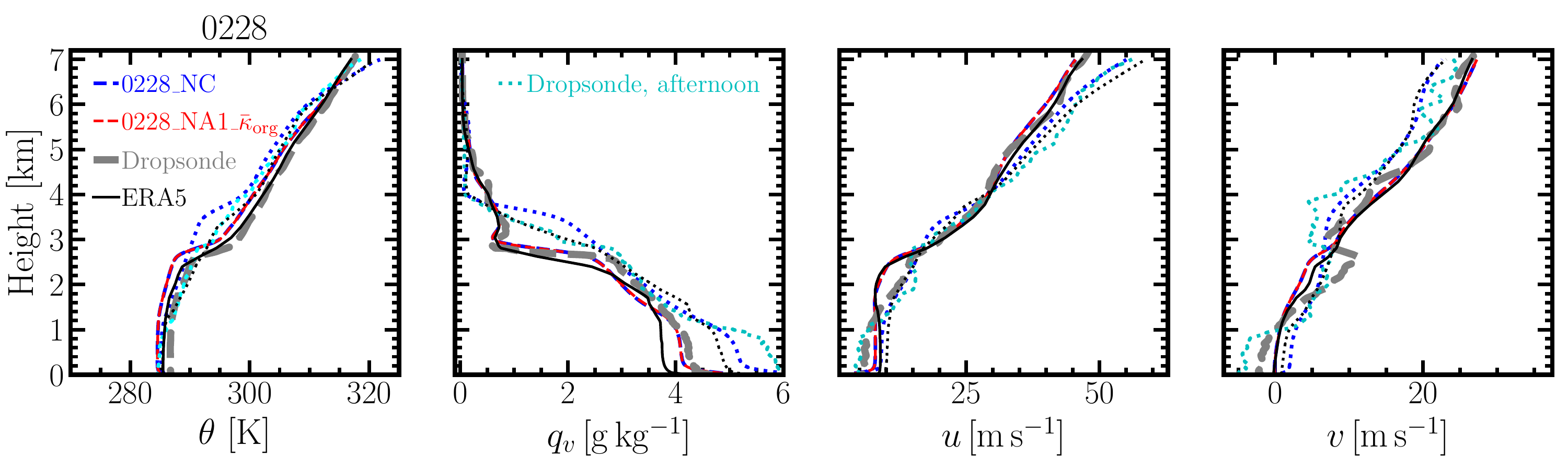}
\includegraphics[width=\textwidth]{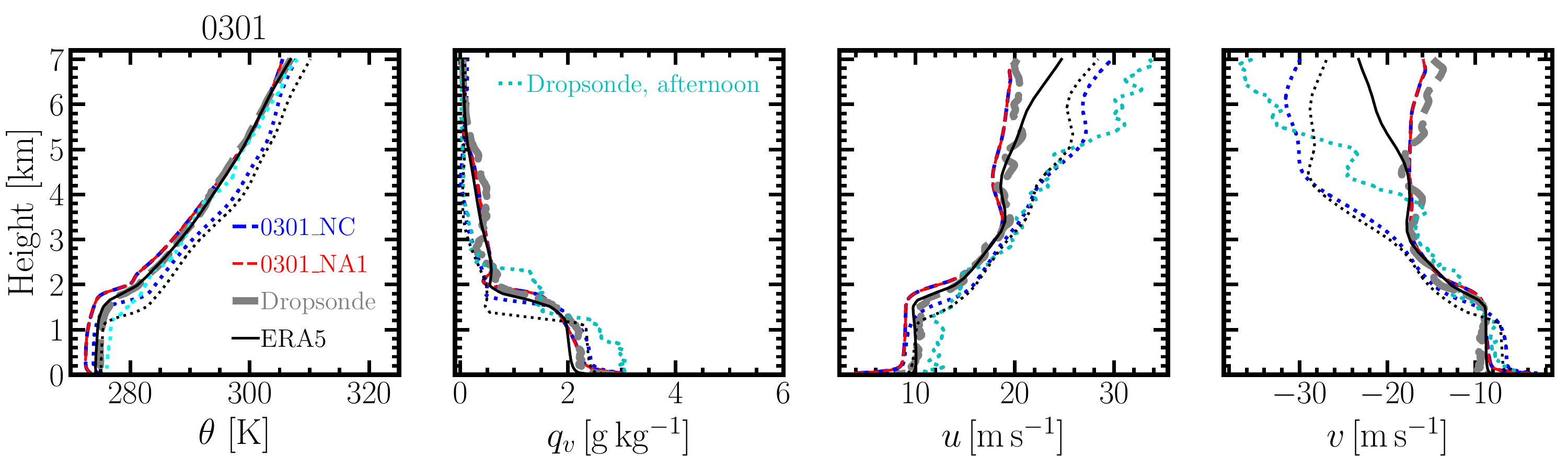}
\end{center}\caption{Comparison of vertical profiles among dropsonde measurements
(grey dashed and cyan dotted lines, averaged over all dropsondes during the morning flight), WRF-LES (blue and red dashed lines, domain averaged), and ERA5 reanalysis (black solid lines, dropsonde center of the morning flight)
averaged during the morning measurement time for the February 28 (upper panel) and March 1 (lower panel) cases.
Dotted lines (with the corresponding colors for ERA5 and LES) represent the ones for the afternoon flights
and only vertical profiles of NC simulations are shown.
11 and 2 dropsondes are released for the morning and
afternoon flights, respectively. The dropsondes location of the morning and afternoon flights are slightly different.
Therefore, the vertical profiles between the morning and
afternoon flights are not quantitatively comparable.
Same simulations as in \Fig{vp_diff}.
See \Tab{tab:runs} for the details of the simulations.
}
\label{verticalP_comp}
\end{figure*}

\begin{figure*}[t!]\begin{center}
\includegraphics[width=\textwidth]{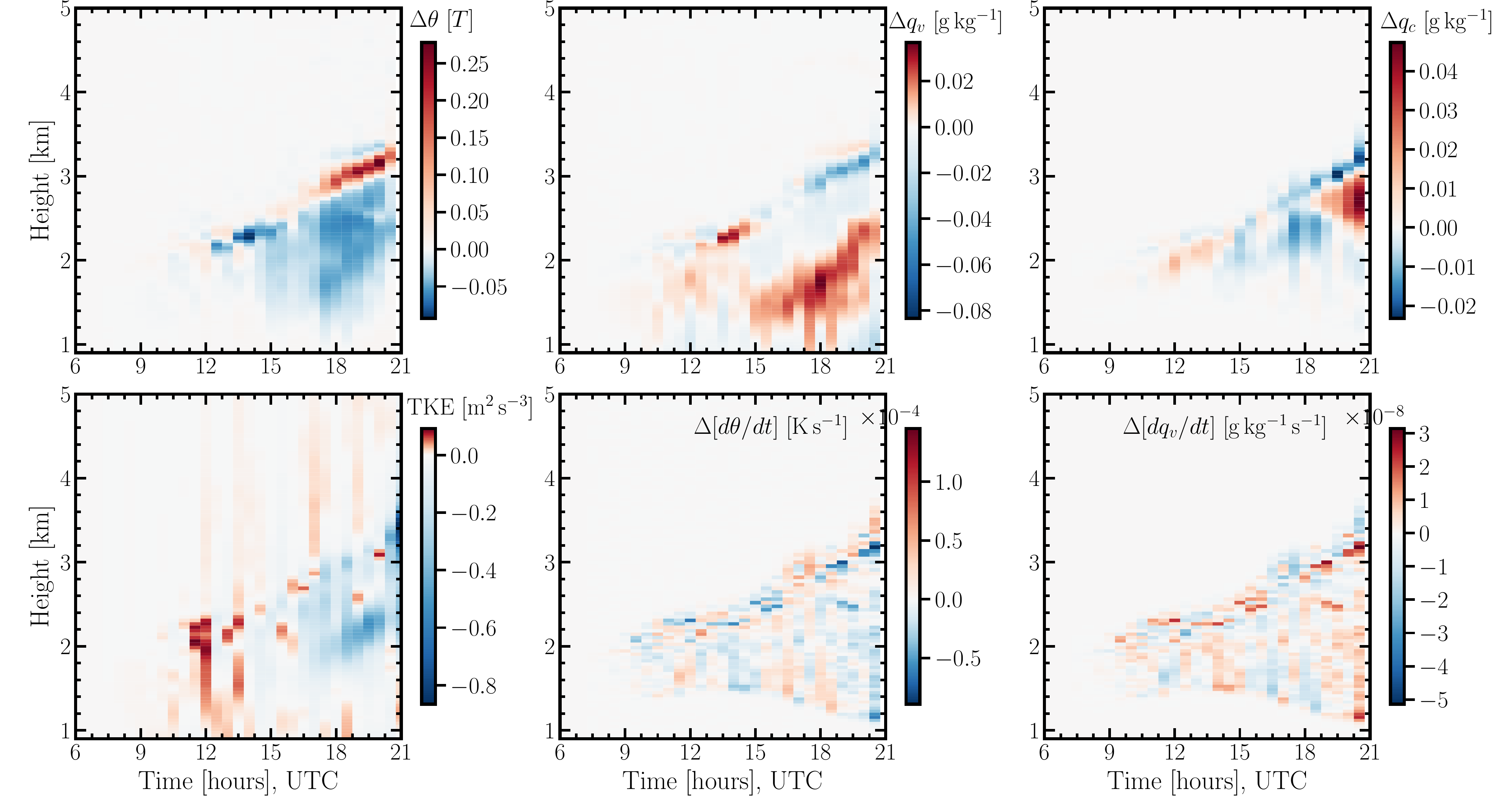}
\end{center}\caption{Evolution of differences of vertical profiles
between simulation $\febm$ and $0228\_{\rm NC}$. Same simulations as in \Fig{vp_diff}.
%\blue{How does $N_a$ affect these quantities?} 
}
\label{vp_t_diff_0228}
\end{figure*}

\begin{figure*}[t!]\begin{center}
\includegraphics[width=\textwidth]{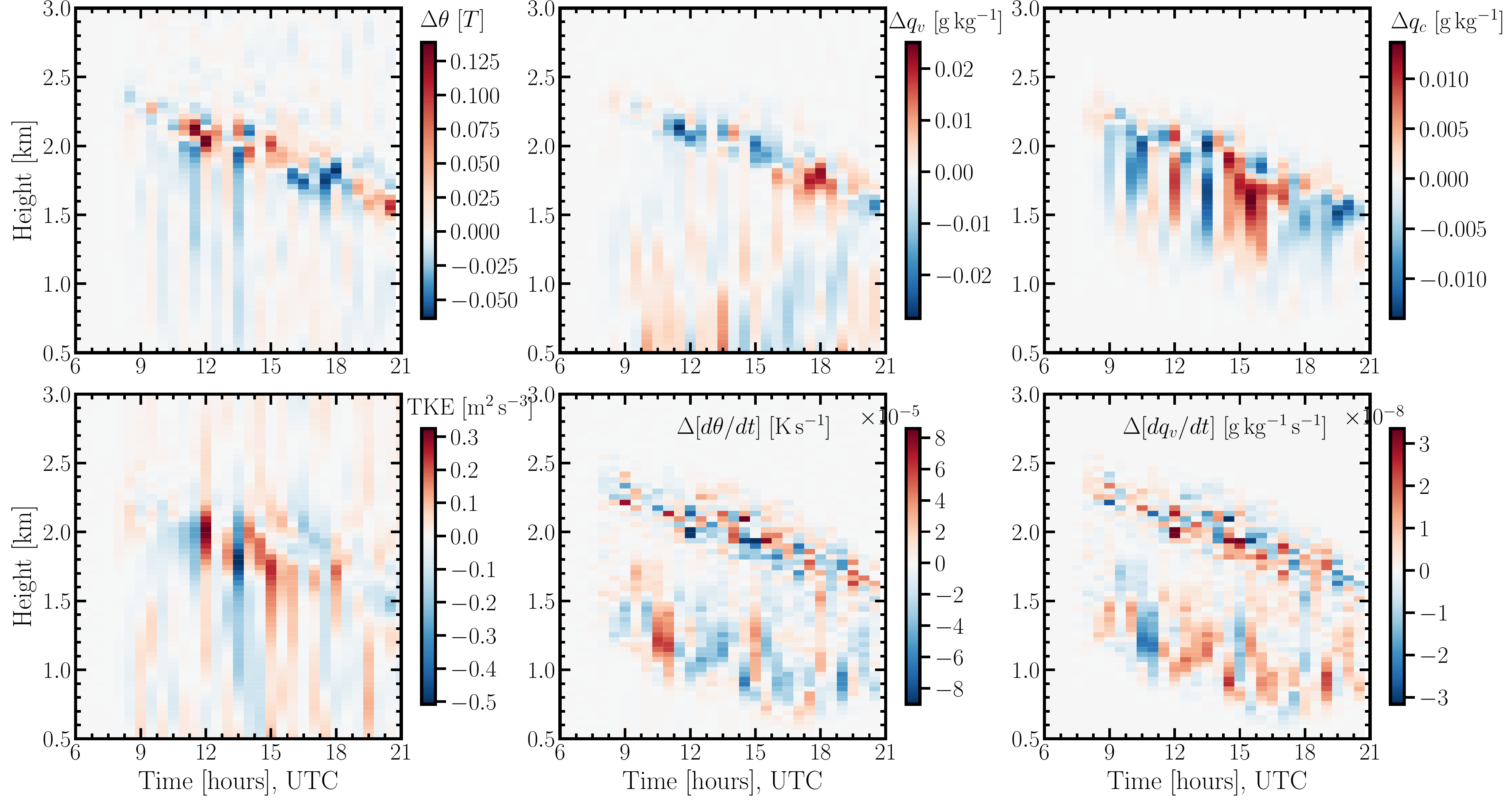}
\end{center}\caption{Evolution of differences of vertical profiles
between simulation $0301\_{\rm NA}1$ and $0301\_{\rm NC}$. Same simulations as in \Fig{vp_diff}.
}
\label{vp_t_diff_0301}
\end{figure*}

\Fig{verticalP_comp} shows the vertical profiles of $\theta$,
$q_v$, $q_c$, $u$, $v$, and $w$ for the February 28 and March 1 cases.
Vertical profiles from WRF-LES
reproduce the observed and reanalysis profiles well for both cases.
The dropsonde measurements for the two CAO cases during the ACTIVATE
campaign are described in \citet{2021arXiv210706193L}.
\Fig{vp_t_diff_0228} and \Fig{vp_t_diff_0301} show the evolution of
differences of vertical profiles between simulation $\febm$ and $0228\_{\rm NC}$ and between simulation $0301\_{\rm NA}1$ and $0301\_{\rm NC}$, respectively.

\section{Retrieve $N_c$ and $r_{\rm eff}$ from GOES-16}
\label{app:Nc}

GOES-16 $N_c$ is derived from cloud effective radius in $\mu$m
and cloud optical depth $\tau$ under the adiabatic assumption \citep{painemal2011assessment},
\EQ
\label{eq:Nc}
N_c = \Gamma^{1/2}\frac{10^{1/2}}{4\pi\rho_w^{1/2}k}\frac{\tau^{1/2}}{r_{\rm eff}^{5/2}},
\EN
where $\Gamma$ ($\rm{g\, m}^{-4}$) is the
the lapse rate due to condensation of water vapor and
is estimated from the cloud top temperature and pressure retrievals
of GOES-16. $\rho_w=1000\, \rm{kg\, m}^{-3}$ is the density of water.
$k = r_v^3/r_{\rm eff}^3=0.8$ is assumed to be a constant with $r_v$ the volume
mean radius. $N_c$ is assumed to be height independent in \Eq{eq:Nc}.

\clearpage

\end{document}